\documentclass[useAMS,usenatbib]{mn2e}
\usepackage{amsmath}

\newcommand {\hi} {{\rm H}\,{\small\rm I}}

\newcommand {\kpc} {\,{\rm kpc}}

\newcommand {\mo}{\,{M}_\odot}

\newcommand{\Gyr}{\,{\rm Gyr}}
\newcommand{\K}{\,{\rm K}}
\newcommand{\mopc}{\,M_{\odot} {\rm pc}^{-2}}
\newcommand {\moyr}{\,M_{\odot} {\rm yr}^{-1}}
\newcommand {\moyrkpc}{M_\odot {\rm yr}^{-1} {\rm kpc}^{-2}}
\newcommand {\moyrmpd}{M_\odot {\rm yr}^{-1} {\rm Mpc}^{-3}}
\newcommand{\gsim}{\lower.7ex\hbox{$\;\stackrel{\textstyle>}{\sim}\;$}}
\newcommand{\lsim}{\lower.7ex\hbox{$\;\stackrel{\textstyle<}{\sim}\;$}}

\newcommand {\apj}{ApJ}
\newcommand {\aj}{AJ}
\newcommand {\apjs}{ApJS}
\newcommand {\apjl}{ApJL}
\newcommand {\mnras}{MNRAS}
\newcommand {\aap}{A\&A}
\newcommand {\aapr}{A\&ARv}
\newcommand {\araa}{ARA\&A}
\newcommand {\pasj}{PASJ}
\newcommand {\pasp}{PASP}

\newcommand {\fcp}{Fundamentals of Cosmic Physics}
\newcommand {\nat}{Nature}
\usepackage{ifpdf}
 \ifpdf
      \usepackage[pdftex]{epsfig}
    \else
      \usepackage[dvips]{epsfig}
\fi
\usepackage{float}

\title[Gas accretion using the Kennicutt-Schmidt law]{Estimating gas accretion in disc galaxies using the Kennicutt-Schmidt law}

\author[Filippo Fraternali \& Matteo Tomassetti]{Filippo
  Fraternali$^{1, 2}$\thanks{E-mail:
    filippo.fraternali@unibo.it} 
and Matteo Tomassetti$^{1,3,4}$\thanks{E-mail: mtomas@astro.uni-bonn.de}\\
$^{1}$Astronomy Department, University of Bologna, via Ranzani 1,
40127, Bologna (IT)\\
$^{2}$Kapteyn Astronomical Institute, Postbus 800, 9700 AV, Groningen
(NL)\\
$^{3}$Argelander~ Institut ~f\mbox{\"{u}}r~ Astronomie,
University~of~Bonn, Auf ~dem ~H\mbox{\"{u}}gel ~71, 53121, Bonn (D)\\ 
$^{4}$Max~ Planck ~Institut~f\mbox{\"{u}}r~ Radioastronomie,
 Auf ~dem ~H\mbox{\"{u}}gel ~69, 53121, Bonn (D)\\ 
}

\begin{document}

\date{Accepted 2012 Month day. Received 2012 Month day}

\pagerange{\pageref{firstpage}--\pageref{lastpage}} \pubyear{2011}

\maketitle

\label{firstpage}

\begin{abstract}

We show how the existence of a relation between the star formation rate and
the gas density, i.e.\ the Kennicutt-Schmidt law, implies a
continuous accretion of fresh gas from the environment into the discs of spiral galaxies.
We present a method to derive the gas infall rate in a galaxy disc as a function of time and radius, and we apply it to the disc of the Milky Way and 21 galaxies from the THINGS sample.
For the Milky Way, we found that the ratio between the past and current star formation rates is about $2-3$, averaged over the disc, but it varies substantially with radius.
In the other disc galaxies there is a clear dependency of this ratio with galaxy stellar mass and Hubble type, with more constant star formation histories for small galaxies of later type.
The gas accretion rate follows very closely the SFR for every galaxy and it dominates the evolution of these systems.
The Milky Way has formed two thirds of its stars after $z=1$, whilst the mass of cold gas in the disc has remained fairly constant with time.
In general, all discs have accreted a significant fraction of their gas after $z=1$.
Accretion moves from the inner regions of the disc to the outer parts, and as a consequence star formation moves inside-out as well.
At $z=0$ the peak of gas accretion in the Galaxy is at about $6-7 \kpc$ from the centre.

\end{abstract}


\section{Introduction}

Star formation is the fundamental process that shapes galaxies into different classes.
Although the majority of stars in the local Universe are found in spheroidal
systems, most of the star formation is contributed by disc
galaxies of the later types (beyond Sb).
The key ingredient for star formation is cold ({\it star-forming})
gas, which is present almost exclusively in disc galaxies. 
However, the amount of cold gas currently available in galaxy discs
appears rather scant.
\citet{Kennicutt98a} estimated that disc galaxies have current star
formation rates ranging from a few to about $\simeq 10 \moyr$.  
Thus, considering a typical gaseous mass of a few $10^9 \mo$, the gas
consumption time scale (i.e.\ the time needed to exhaust the gas fuel
with a constant star formation rate) is always of the order of a few
Gigayears.
This result, known as the gas-consumption dilemma \citep{Kennicutt83}, suggests the need
for continuous accretion of cold gas onto galaxy discs \citep[e.g.][]{Sancisi+08}.

Several pieces of evidence show that disc galaxies should 
collect fresh gas from the environment in order to sustain their star
formation. 
In the Milky Way, the star formation rate (SFR) in the solar
neighbourhood appears to have remained rather constant in the last
$\sim10 \Gyr$ \citep[e.g.][]{Twarog80, Rocha-Pinto+00, Binney+00}, suggesting
a continuous replenishment of the gas supply. 
Simple (closed-box) models of chemical evolution for our Galaxy
predict too few metal deficient \textit{G-dwarf} stars than observed
\citep{Searle&Sargent72, Pagel&Patchett75, Haywood01}. 
These observations are easily explained by accounting for infall of
fresh unpolluted gas \citep[e.g.][]{Chiappini+97, Chiappini+01}.
Observations of Damped Lyman Alpha systems show almost no evolution in
the neutral gas content of structures in the Universe
\citep[][]{Zwaan+05,Lah+07,Prochaska+09}. 
\citet{Hopkins+08} pointed out that this constancy of gas density can
be explained assuming a rate of gas replenishment proportional to the
universal SFR density.
\citet{Bauermeister+10} converged to a similar result when comparing
the evolution of the molecular gas depletion rate with that of the
cosmic star formation history (SFH).
Finally, the derivations of SFHs for galaxies with different stellar
masses consistently show that late type systems do have a rather
constant SFR throughout the whole Hubble time \citep[e.g.][]{Panter+07}.
These findings, other than being the signature of {\it downsizing} in
cosmic structures, point at a continuous infall of gas onto galaxies of
late Hubble types.

The way gas accretion into galaxies takes place is still a matter of debate.
The classical picture states that thermal instabilities should cause
the cooling of the hot coronae that surround disc galaxies and be the
source of cold gas infall \citep[e.g.][]{Maller&Bullock04, Kaufmann+06}.
However, recent studies have shown that, due to a combination of
buoyancy and thermal conduction, hot coronae turn out to be remarkably stable
and thermal instability does not appear to be a viable mechanism for 
gas accretion \citep{Binney+09, Joung+11}.
On the other hand, accretion may take place in the form of cold flows
but the importance of this process is expected to drop
significantly for redshift $z<2$ \citep{Dekel&Birnboim06, vdVoort+11}. 
Observations of local gas accretion at
21-cm emission seem to show too little gas around
galaxies in the form of \hi\ (high-velocity) clouds 
to justify an efficient feeding of the disc star formation
\citep{Sancisi+08, Thom+08, Fraternali09}.
However, UV absorption towards quasars and halo stars point to the
possibility that ionised gas at temperatures
between a few 10$^4$ and a few $10^5 \K$ could fill the gap between expectations
and data \citep{Bland-Hawthorn09, Collins+09, Lehner&Howk11}. 

In this paper, we estimate gas accretion in galactic discs indirectly
by comparing basic physical properties of galaxies today.
Our model relies on the existence of a law relating the star formation
rate density (SFRD) and the gas surface density ($\Sigma_{\rm gas}$)
holding at every redshift, i.e.\ the Kennicutt-Schmidt (K-S) law
\citep{Schmidt59, Kennicutt98a}.
This approach has similarities with the model of \citet{Naab&Ostriker06} that we describe in detail in Section \ref{sec:discussion}.
The paper is organized as follows. 
Section \ref{sec:model} describes our method. In Section
\ref{sec:data} we apply our model to the Milky Way disc and to a
sample of external discs. 
In Section \ref{sec:discussion} we discuss our results.  
Section \ref{sec:conclusions} sums up.

\section{Description of the Model}\label{sec:model}

The evolution of gas and stellar densities ($\Sigma_{\rm gas}$ and $\Sigma_{\rm *}$) in a galactic disc as a function of the lookback time $t$ is described by the following equations \citep{Tinsley80}:
\begin{align}
\label{eq:tinsley1}
&-\dfrac{{\rm d}\Sigma_{\rm *}}{\rm dt}=+{\rm SFRD}-\dot\Sigma_{\rm fb}
\\
\label{eq:tinsley2}
&-\dfrac{{\rm d}\Sigma_{\rm gas}}{\rm dt}=-{\rm SFRD}+\dot\Sigma_{\rm fb}+\dot\Sigma_{\rm ext}
\end{align}
where ${\rm SFRD}$ is the star formation rate density, $\dot\Sigma_{\rm fb}$ is the contribution from stellar feedback and $\dot\Sigma_{\rm ext}$ is the inflow/outflow rate from/to the external environment.
${\rm SFRD}$ gives the rate at which stars are formed or, with a sign inversion, the rate at which gas is consumed, whilst $\dot\Sigma_{\rm fb}$ represents the gas density per unit time returned by stellar evolution to the gas reservoir in the disc.
We treat the stellar feedback with the instantaneous recycling approximation (I.R.A.), meaning that the fractional mass that stars return to the ISM at each time is assumed to be constant and equal to the return factor $\mathcal{R}$.
Equations (\ref{eq:tinsley1})-(\ref{eq:tinsley2}) modify as follows:
\begin{align}
\label{eq:tinsley1wR}
&-\dfrac{{\rm d}\Sigma_{\rm *}}{\rm dt}=(1-\mathcal{R}){\rm SFRD} \\
\label{eq:tinsley2wR}
&-\dfrac{{\rm d}\Sigma_{\rm gas}}{\rm dt}=-(1-\mathcal{R}){\rm SFRD}+\dot\Sigma_{\rm ext}
\end{align}
In Section \ref{sec:nonIRA} we discuss the effect of a delayed return.

The parameter $\mathcal{R}$ in eqs.\ (\ref{eq:tinsley1wR})-(\ref{eq:tinsley2wR}) depends on the IMF and the fraction $\eta(M)$ of mass that a star of mass $M$ returns to the ISM, called the \textit{stellar initial versus final mass relation}. 
Following the prescription of \citet{Kennicutt+94} we assume that high-mass stars ($M > 8$ M$_\odot$) all leave a $1.4$ M$_\odot$ remnant. 
Low-mass stars are assumed to leave remnants with masses between $\sim0.5$ and $\sim1.3$ M$_\odot$, with a linear dependency from the initial mass \citep[see Chapter 2 of][for details]{Matteucci01}.
Assuming a Salpeter IMF, our fiducial value for the recycling parameter is $\mathcal{R} \simeq 0.30$.
Assuming Kroupa \citep{Kroupa+93} or Chabrier \citep{Chabrier03} IMFs would make little difference, giving $\mathcal{R}=0.31$ and $\mathcal{R}=0.32$ (or $\mathcal{R}=0.46$ for a flatter high-mass slope) respectively. 
With this value the gas consumption timescales are extended by a factor $(1-\mathcal{R})^{-1}=1.4$ \citep[see][]{Kennicutt+94}.

The third term on the RHS of eq.\ (\ref{eq:tinsley2}) is the inflow/outflow term. 
The combination of eqs.\ (\ref{eq:tinsley1}) and (\ref{eq:tinsley2}) leads to an expression for this term:
\begin{equation}\label{acc_prima}
\dot\Sigma_{\rm ext}=-\dfrac{{\rm d}\Sigma_{\rm *}}{\rm dt}-\dfrac{{\rm d}\Sigma_{\rm gas}}{\rm dt}
\end{equation}
In the following we often refer to this term as the accretion (or infall) rate.

Equations (\ref{eq:tinsley1}) and (\ref{eq:tinsley2}) can be applied to a galactic disc assuming that they are valid at each radius, i.e.\ that the disc can be divided into annuli that evolve independently.
This assumption is equivalent to say that stars that form at a certain radius remain at that radius for the lifetime of the galactic disc.
In Section \ref{sec:migration} we discuss the effect of stellar migration and show that our main results do not change significantly.
As a consequence of this assumption we can integrate eq.\ (\ref{eq:tinsley1}) from $t_{\rm form}$ (the lookback time at the disc formation) to $t=0$ to get the current stellar surface density.

Clearly, in order to solve the above system one needs the ${\rm SFRD}$ as a function of lookback time and radius, i.e.\ the SFH as a function of radius. 
In the Milky Way, the SFH is known with fairly good precision only in the Solar Neighborhood \citep[e.g.][]{Rocha-Pinto+00, Cignoni+06}. 
In external galaxies there have been pioneering attempts to estimate the SFH as a function of galactic radius \citep[e.g.][]{Gogarten+09, Weisz+11}.
In general, however, a precise determination of ${\rm SFH}(R)$ is beyond the capability of the current data.
Therefore, for the time being, we must content ourselves with using simple parametrisations for the shape of the SFH.
In the following section we describe how we estimate the trend of the SFH at each radius in a galactic disc.
We consider only the disc, leaving aside the bulge/bar, which may have formed and evolved differently as it shows a markedly different SFH \citep[e.g.][]{Hopkins+01}.

\subsection{Reconstructing the SFH of a galaxy disc}
\label{sec:sfh}

We assume that the SFH in a galaxy disc can be described by a dimensionless and positive function $f=f(R,t)$.
This allows us to write the star formation rate density as:
\begin{equation}
\label{eq:sfrd}
{\rm SFRD}(R,t) = {\rm SFRD}(R,0) \times f(R,t)
\end{equation}
where ${\rm SFRD}(R,0)$ is the radial distribution of the current star formation rate density, which we derive from the data.
The simplest expression for $f$ one can assume is a first order polynomial with time:
\begin{equation}\label{eq:f1}
f_1(R,t) = 1 + \left[\gamma(R)-1\right] \frac{t}{t_{\rm form}}  \quad (t_{\rm form}\geq t \geq 0)
\end{equation}
where $t$ is the lookback time and $t_{\rm form}$ is the age of the stellar disc.
With this definition, $\gamma(R)$ is the value of the function $f$ at $t=t_{\rm form}$ and sets the steepness of the SFH.
A value of $\gamma(R)$ equal to unity represents a constant SFR over the disc lifetime, $\gamma(R)>(<)1$ describes a SFR that increases (decreases) with increasing lookback time.
\footnote{Note that $\gamma(R)<0$ would imply a lower formation time at that radius; this occurance does not arise with our choice of parameters.}
The dependence of $\gamma$ on the radius $R$ allows us to describe different shapes of SFH within the disc.
In Section \ref{sec:fs} we discuss the effect of taking different forms for $f$, we anticipate that our main results remain unchanged.

We define the \textit{global} $\gamma$ for a galactic disc as:
\begin{equation}\label{eq:gammaDef}
\gamma = \frac{2 \pi \int_0^{R_{\rm m}} R\, \gamma(R)\,{\rm SFRD}(R,0)\,{\rm d}R}{2 \pi \int_0^{R_{\rm m}} R\, {\rm SFRD}(R,0)\,{\rm d}R}
\end{equation}
where the integration is performed out to the maximum radius $R_{\rm m}$; the denominator is just the current SFR of the disc.
This definition allows us to write an expression for the global SFR analogous to eqs.\ (\ref{eq:sfrd}) and (\ref{eq:f1}),
\begin{equation}\label{eq:sfr}
{\rm SFR}(t) = {\rm SFR}(0) \times \left[1+(\gamma-1)\frac{t}{t_{\rm form}}\right]
\end{equation}

\begin{figure}
\centering
\includegraphics[width=0.5\textwidth]{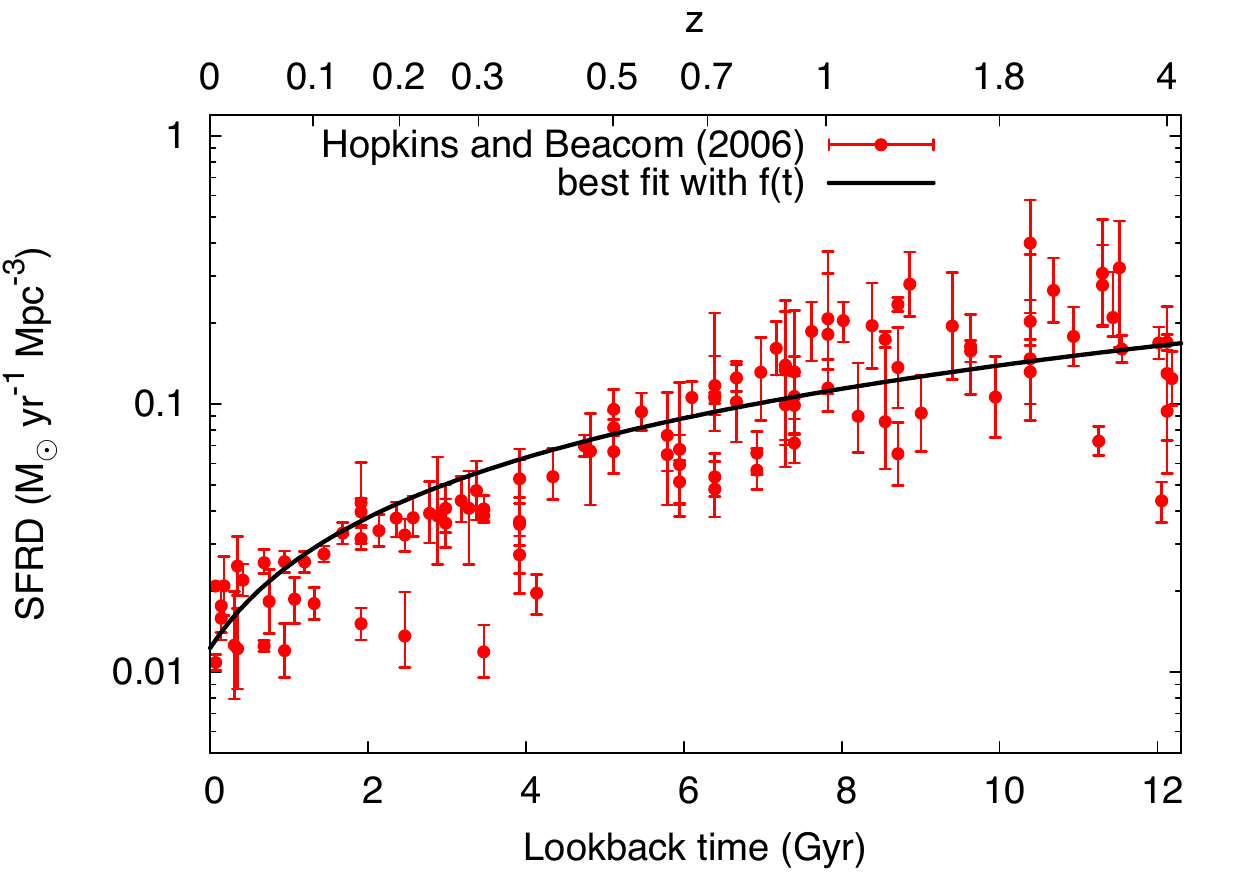}
\caption{\label{fig:SFHUniverse}
Evolution of the average SFR density of the Universe, the points are taken from \citet{Hopkins&Beacom06}.
The thick black curve shows a fit with our function $f(t)$ described in the text.
}
\end{figure}

As an illustration of the meaning of our parameter $\gamma$ in Fig.\ \ref{fig:SFHUniverse} we fit the average SFR density of the Universe with eq.\ (\ref{eq:sfr}).
The data points are taken from \citet{Hopkins&Beacom06}.
Here we take $t_{\rm form} = 12$ Gyr.
For the other parameters we infer $\rm{SFR}(0)=0.012 \moyrmpd$ and $\gamma = 13.4$ as the best-fit values.
This fit shows that the SFR density of the Universe as a whole is described by a value of $\gamma$ much larger than unity, as expected given that the SFR density in the past was much higher than now. 
The shape of the universal SFR density is produced by the combined contribution of galaxies with very different SFHs.
Massive red-sequence galaxies dominate the overall SFR at high-z, whilst late-type galaxies become relatively more and more important as time passes \citep[e.g.][]{Panter+07, Vincoletto+12}.
Although our parametrisation is only suitable for galaxies with non-negligible current SFR, taking $\gamma$ as the ratio of ${\rm SFR(t_{\rm form})}/{\rm SFR(0)}$, clearly red sequence galaxies must have $\gamma \gg 1$.

The above assumptions for the SFRD (eqs.\ \ref{eq:sfrd} and \ref{eq:f1}) allow us to integrate eq.\ (\ref{eq:tinsley1wR}) from $t_{\rm form}$ to 0 to obtain:
\begin{equation}\label{eq:gammaR}
\gamma(R)=\frac{2~\Sigma_*(R,0)}{(1-\mathcal{R})~t_{\rm form}{\rm SFRD}(R,0)}-1
\end{equation}
where $\Sigma_*(R,0)$ is the surface density profile of the stellar disc that we derive from the observed surface brightness profile.

Eq.\ \ref{eq:gammaR} reveals that our parameter $\gamma(R)$ is ultimately a ratio between the stellar density and the current SFRD, given a certain formation time for the disc, which is of the order of the Hubble time and can be assumed roughly equal for all discs.
A $\gamma(R)$ close to 1 means that the current SFR density extended to an Hubble time produces as many stars as observed at that location in the disc.
Integrating $\gamma(R)$ over the disc and making used of eq.\ (\ref{eq:gammaDef}) we find an expression for the global $\gamma$ as:
\begin{equation}\label{eq:gamma}
\gamma=\frac{2~M_*(0)}{(1-\mathcal{R})~t_{\rm form}{\rm SFR}(0)}-1
\end{equation}
where $M_*(0)$ and ${\rm SFR}(0)$ are respectively the current stellar mass and star formation rate of the entire disc.

Clearly from eq.\ (\ref{eq:gamma}), one can also describe $\gamma$ as proportional to the ratio between the average past SFR and the current SFR.
This latter is the inverse of the so-called Scalo $b$-parameter \citep{Scalo86} to which $\gamma$ is closely related (see Section \ref{sec:things}).
We also note that any modification of eq.\ (\ref{eq:f1}) with power law terms of $\frac{t}{t_{\rm form}}$ would produce the same dependencies for $\gamma$ as eq.\ (\ref{eq:gammaR}).
We discuss these points further in Section \ref{sec:fs}.

\subsection{The star formation law}
\label{sec:sfLaw}

The main ingredient of our model is the relation between the rate of star formation and the gas density.
The empirical star formation law is broadly accepted to be a power-law relation between the ``cold'' gas density and the star formation rate density \citep{Schmidt59, Kennicutt98a}.
This relation has been tested on global (i.e. averaging gas and star formation on the whole disc) and local scales \citep{Kennicutt+07} considering the molecular gas or the sum of atomic and molecular gas.
Remarkably, it appears to hold across various orders of magnitudes, from quiescent discs to star-busting galaxies \citep{Kennicutt98b, Krumholz+11}.
The dependence on gas metallicity is difficult to investigate due to the related dependency of the X$_{\rm CO}$ factor \citep{Boissier+03}.
However, the compilation of azimuthal averages of SFRD and $\Sigma_{\rm gas}$ presented by \citet{Leroy+08} for the THINGS sample, which includes galaxies with very different masses and presumably different metallicities, seems to show that the effect is limited \citep[see][and Section \ref{sec:Others}]{Wyder+09}.
Thus, in the following we assume the following star formation law:
\begin{equation}\label{eq:sfLaw}
{\rm SFRD}=A\Sigma_{\rm gas}^N
\end{equation}
with $N=1.4$ and $A=1.6 \times 10^{-4}$ if $\Sigma_{\rm gas}$ is measured in $\mopc$ and SFRD in $\moyrkpc$ \citep{Kennicutt98a}.
Note that the above value of $A$ takes into account the correction for the presence of Helium (a factor 1.36 in mass), not applied in \citet{Kennicutt98a}.

Deviations from the above relation are observed at column densities lower than $\Sigma_{\rm gas,\, th} \simeq 10 \mopc$, referred to as the density threshold. 
Below these densities the law steepens and the scatter increases making it more difficult to describe \citep[][]{Schaye04, Bigiel08}.
In the context of this paper, it may have an effect in the outer parts of the discs (see Sections \ref{sec:MW} and \ref{sec:Others}). 

\citet{Krumholz+12} have recently pointed out that
the star formation law expressed in terms of  volume density is
more general than the one expressed in terms of  surface density.
This is very relevant to understand the physical origin of the law and to perform 
numerical simulations. 
However, for our investigation a surface-density law is far more practical. 
Clearly, the choices are equivalent if in the region of interest, i.e.\ within the star forming disc, the scale-height of the gaseous disc does not change much as a function of radius. 
The validity of this condition (within a factor 2) is well established in our Galaxy, as the disc appears to be flaring significantly only beyond $R \simeq 16 \kpc$ \citep{Kalberla&Dedes08}.

\subsection{Inferring the gas accretion rate}

The above assumptions on the SFH and the star formation law allow us to infer the local gas accretion rate using eq. (\ref{acc_prima}).
If the star formation law is described by eq.\ (\ref{eq:sfLaw}), the accretion rate density can be written as:
\begin{eqnarray}\label{eq:gar}
\dot\Sigma_{\rm ext}(R,t) & = & (1-\mathcal{R}){\rm SFRD}(R,0)f(R,t)  \nonumber\\
& & -\frac{{\rm SFRD}^{1/N}(R,0)}{N\, A^{1/N}\,t_{\rm form}}  f(R,t)^\frac{1-N}{N}\frac{{\partial f}(R,t)}{\partial t}
\end{eqnarray}
where ${\rm SFRD}(R,0)$ is the current total gas density, $A$ and $N$ are the parameters of the star formation law and $f$ is the functional form of the star formation history.
The RHS of eq.\ (\ref{eq:gar}) can also be expressed in terms of $\Sigma_{\rm gas}(R,0)$ (using the K-S law) instead of ${\rm SFRD}(R,0)$.
This equation is general and it could be used if the {\it exact} shape of the ${\rm SFH}(R)$ were known.
If $f$ is given by eq.\ (\ref{eq:f1}), the gas accretion rate density becomes:
\begin{eqnarray}\label{eq:garf1_1}
\lefteqn{\dot\Sigma_{\rm ext}(R,t)= (1-\mathcal{R}){\rm SFRD}(R,0)\left[1+(\gamma(R)-1)\frac{t}{t_{\rm form}}\right]  } \nonumber \\
& & -\frac{{\rm SFRD}^{1/N}(R,0)}{N\, A^{1/N}\,t_{\rm form}} \left(\gamma(R)-1\right) \left[1+\left(\gamma(R)-1\right)\frac{t}{t_{\rm form}}\right]^\frac{1-N}{N}
\end{eqnarray}
or, equivalently in terms of the gas density:
\begin{eqnarray}\label{eq:garf1_2}
\lefteqn{\dot\Sigma_{\rm ext}(R,t)= A(1-\mathcal{R})\Sigma_{\rm gas}^N(R,0) \left[1+(\gamma(R)-1)\frac{t}{t_{\rm form}}\right]} \nonumber \\
& & -\frac{\Sigma_{\rm gas}(R,0)}{N\, t_{\rm form}} \left(\gamma(R)-1\right) \left[1+\left(\gamma(R)-1\right)\frac{t}{t_{\rm form}}\right]^\frac{1-N}{N}
\end{eqnarray}

The above equations give the rate at which fresh gas must be added to the disc at a certain radius and as a function of time to produce the star formation density given by ${\rm SFRD}(R,t)$.
As shown above, the shape of the SFH is regulated by the value of $\gamma(R)$.
For $\gamma(R)= 1$, SFR and accretion are constant in time. 
More in general, when $\gamma$ is close to 1 the second term on the RHS of eq.\ (\ref{eq:garf1_1}) is small leading to $\dot\Sigma_{\rm ext}(R,t) \simeq (1-\mathcal{R}){\rm SFRD}(R,t)$.
Therefore the gas needed is directly proportional to the gas consumed by the star formation and this proportionality is simply $(1-\mathcal{R})$.
For $\gamma(R) > 1$ the second term in eqs.\ (\ref{eq:garf1_1}) and (\ref{eq:garf1_2}) becomes gradually more negative and less accretion is needed.
High values of $\gamma(R)$ may result in a negative $\dot\Sigma_{\rm ext}(R,t)$, hence at that time and radius, the amount of gas available in the disc exceeds what is needed by star formation.
This would be the signature of an outflow of gas.

We define the global accretion rate $\dot M_{\rm ext}(t)$ as the gas mass accreted per unit time over the whole galactic disc, obtained integrating eqs.\ (\ref{eq:garf1_1}) or (\ref{eq:garf1_2}) over the disc out to our maximum radius $R_{\rm m}$.
It is evident that the global accretion rate will have dependencies similar to those of the local one and the above considerations apply.
In particular, for galaxies with $\gamma$ of order unity, the global infall rate closely follows the $\rm SFR$ at every time $\dot M_{\rm ext}(t) \simeq (1-\mathcal{R}){\rm SFR}(t)$.

It is worth noting that $\dot \Sigma_{\rm ext}(R,t)$ represents the gas surface density that has to be added to (or removed from) the disc per unit time to produce the reconstructed ${\rm SFRD}(R,t)$.
If we are considering a single annulus, $\dot \Sigma_{\rm ext}(R,t)$ refers generically to the external environment including the adjacent annuli.
Therefore, $\dot \Sigma_{\rm ext}(R,t)$ coincides with the accretion (infall) rate at that radius only if there are no radial flows in the gaseous disc.
This is because, with this method we do not trace the location of the gas infall but rather the radius at which such gas turns into stars.
We discuss this issue a bit further later on.
On the other hand, $\dot M_{\rm ext}(t)$ represents the total mass needed to be accreted by the galaxy disc from outside, at the time $t$, to sustain the SFR at the reconstructed rate.
Note however that also in this case the definition of \textit{outside} the disc is merely \textit{beyond} the maximum radius $R_{\rm m}$ assumed for the disc, which we took to be 5 times the scale-length of the stellar disc.

\section{Results}\label{sec:data}

In this Section we apply the method described above to real galaxy discs. 
We impose the following {\it boundary conditions}:
i) the current stellar density profile $\Sigma_*(R,0)$, ii) the current SFR density profile ${\rm SFRD}(R,0)$.
The latter could be in principle derived from the current gas density profile using the star formation law (Section \ref{sec:sfLaw}).
However here we restrict our analysis to cases where the current SFRD is known from independent estimates.
We first consider the Milky Way and then a sample of spiral galaxies with known SFRD profiles.

\subsection{Application to the Milky Way disc}
\label{sec:MW}

\begin{figure}
\centering
\includegraphics[width=0.5\textwidth]{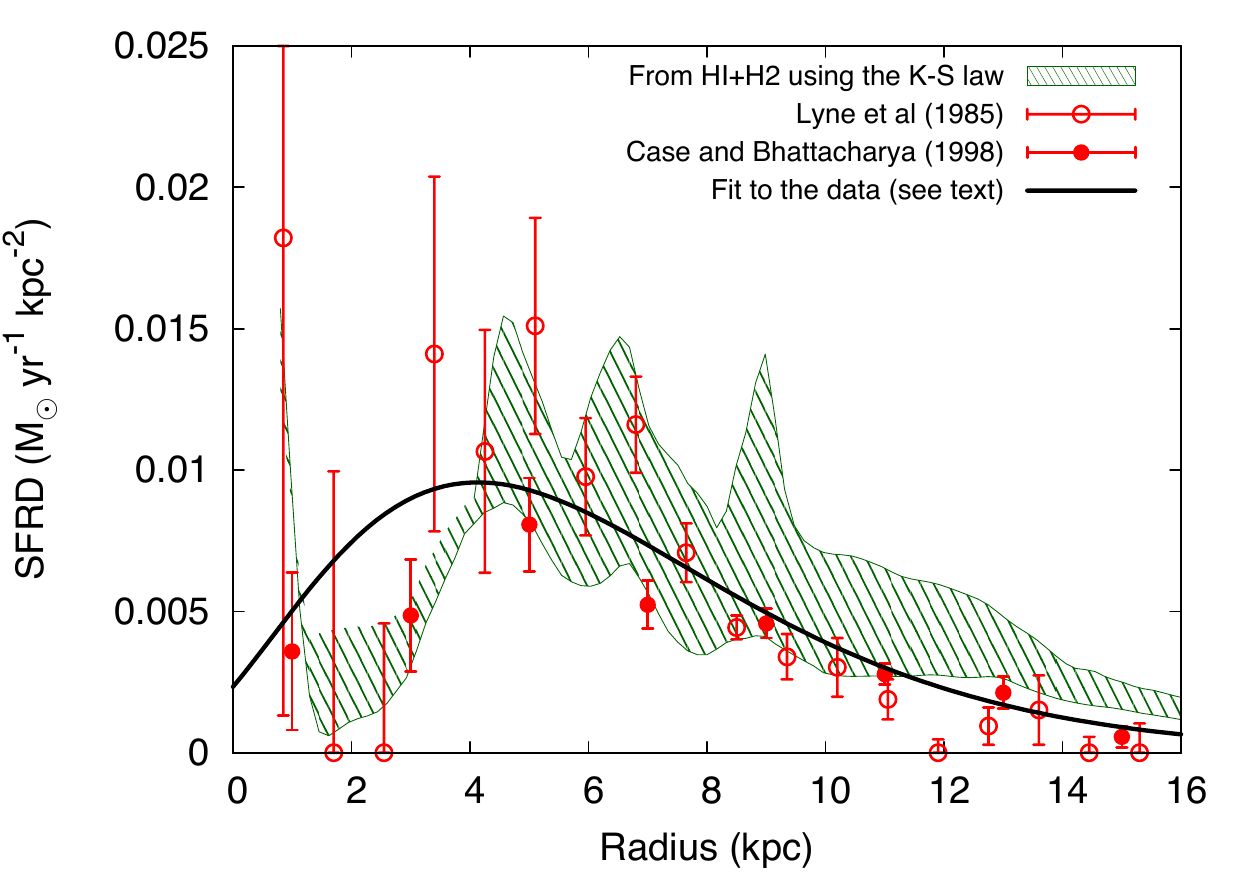}
\caption{\label{fig:sfrdMW}
Current star formation rate density of the Milky Way disc estimated
using pulsars (open squares), and supernova remnants (filled squares).
These data, originally normalized to the Solar neighbourhood value,
have been fitted with a function described in the text (solid line)
and rescaled so that the integral over the disc gives
SFR(0)$=3 \moyr$.
The shaded region shows the ${\rm SFRD}(R,0)$ one would get from the
total (neutral + molecular) gas density using the K-S law.
The lower and upper boundaries are due to the different determinations
of the \hi\ surface densities from \citet{Binney&Merrifield98} and \citet{Kalberla&Dedes08} respectively.
}
\end{figure}

We assume that the stellar disc of the Milky Way is exponential with a
scale length $h_R = 3.2 \kpc$ \citep{Binney&Tremaine08} and a maximum
extension of $R_{\rm m}=5\, h_R$.
The normalization of this profile is chosen so that the total stellar
mass of the disc is $4\times10^{10}$ M$_\odot$
\citep{Dehnen&Binney98}. 
The present distribution of the star formation rate density with respect to the Solar neighborhood, ${\rm SFRD}(R,0)/{\rm SFRD}(R_{\odot},0)$, has been estimated using several methods.
In Fig.\ \ref{fig:sfrdMW} we show the estimates coming from the distribution of pulsars \citep[open squares,][]{Lyne+85} and supernova remnants \citep[filled squares,][]{Case&Bhattacharya98}.
We fit these points with a function (solid line in Fig.\ \ref{fig:sfrdMW}) of the form: 
\begin{equation}
{\rm SFRD}(R,0)={\rm SFRD}(0,0) \left( 1 + \frac{R}{R_{*}} \right) ^{\alpha} e^{-\frac{R}{R_{*}}},
\end{equation}
where $\alpha=3.10$ is an exponent defining the tapering towards the centre
and $R_*=1.96 \kpc$ is a scale radius such that the peak of the distribution is
located at  $R=R_*(\alpha-1)$.
The central SFR density ${\rm SFRD}(0,0)=2.3\times10^{-3} \moyrkpc$ is normalized in order to have a
global current star formation rate of the Milky Way ${\rm SFR}(0)=3
\moyr$ \citep[see Table 1 in][]{Diehl+06}.
The same normalization is then applied to the data points.
The shaded region in Fig.\ \ref{fig:sfrdMW} shows, as a consistency
check, the SFRD inferred from the current total gas density using the
K-S law.
The upper and lower boundaries refer to two different determinations
of the neutral gas density by \citet{Binney&Merrifield98} and
\citet{Kalberla&Dedes08}.
The latter is not determined for $R<4 \kpc$.
The molecular gas density is taken from \citet{Nakanishi&Sofue06}.
On the whole the agreement between the direct determinations of SFRD
and that inferred by the K-S law is very good.
In the outer parts the K-S predicts a larger SFR than observed, likely
due to the low densities and the steepening of the law.
We quantify this effect in Section \ref{sec:Others}.
Some recent determinations of the current SFR of the Milky tend to give values around $1-2 \moyr$ \citep[see e.g.][]{Murray&Rahman10}.
However, Fig.\ \ref{fig:sfrdMW} shows that only a ${\rm SFR}(0) \simeq 3 \moyr$ is self-consistent with the derivation of the ${\rm SFRD}(R,0)$ from the gas surface density by inverting the K-S law, so we prefer to adopt this value throughout the paper.
In the following we further fix the age of the stellar disc $t_{\rm
 form}=10 \Gyr$ or, equivalently, $z_{\rm form}=1.8$.\footnote{We
 assume a standard cosmology with $\Omega_{\rm b}=0.27$ and
 $\Omega_{\rm \Lambda}=0.73$.}
Note that taking a larger value for $t_{\rm form}$, suggested for
instance by the investigation of \citet{Aumer&Binney09} would strengthen
our results.

\begin{figure}
\centering
\includegraphics[width=0.5\textwidth]{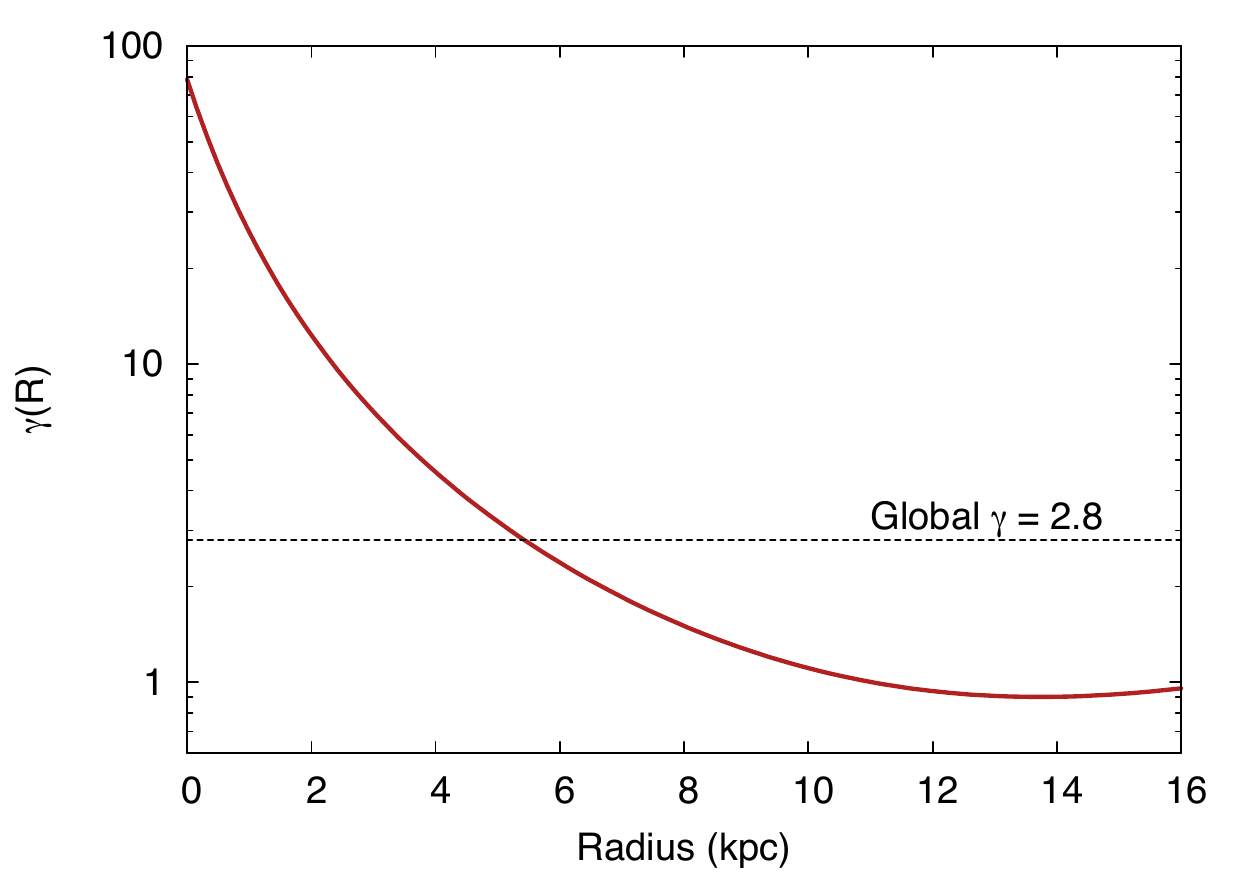}
\caption{\label{fig:gammaR}
Steepness of the SFH as a function of radius for the Milky Way disc, parametrised by $\gamma(R)$ (see Section \ref{sec:sfh}).
The horizontal line shows the global value obtained for the whole disc.
}
\end{figure}

We reconstruct the SFH of the Milky Way disc as described in Section
\ref{sec:sfh}. 
We find a steepness of $\gamma = 2.8$ for the whole disc.
The behavior as a function of radius is shown in Fig.\
\ref{fig:gammaR}. 
The parameter $\gamma(R)$ appears to be a decreasing function of the radius $R$.
Recalling that $\gamma(R)$ is the ratio between the initial and the
current SFRDs, this trend shows, as expected, that star formation in the central
regions must have been faster than in the outer disc, and it has spread from inside out
 \citep[see also][]{Naab&Ostriker06}.
Note that our approach is fully axi-symmetric and it does not account for
the presence of a bar, thus our results for $R\lsim 3 \kpc$ should be
taken with some caution.

\begin{figure}
\centering
\includegraphics[width=0.5\textwidth]{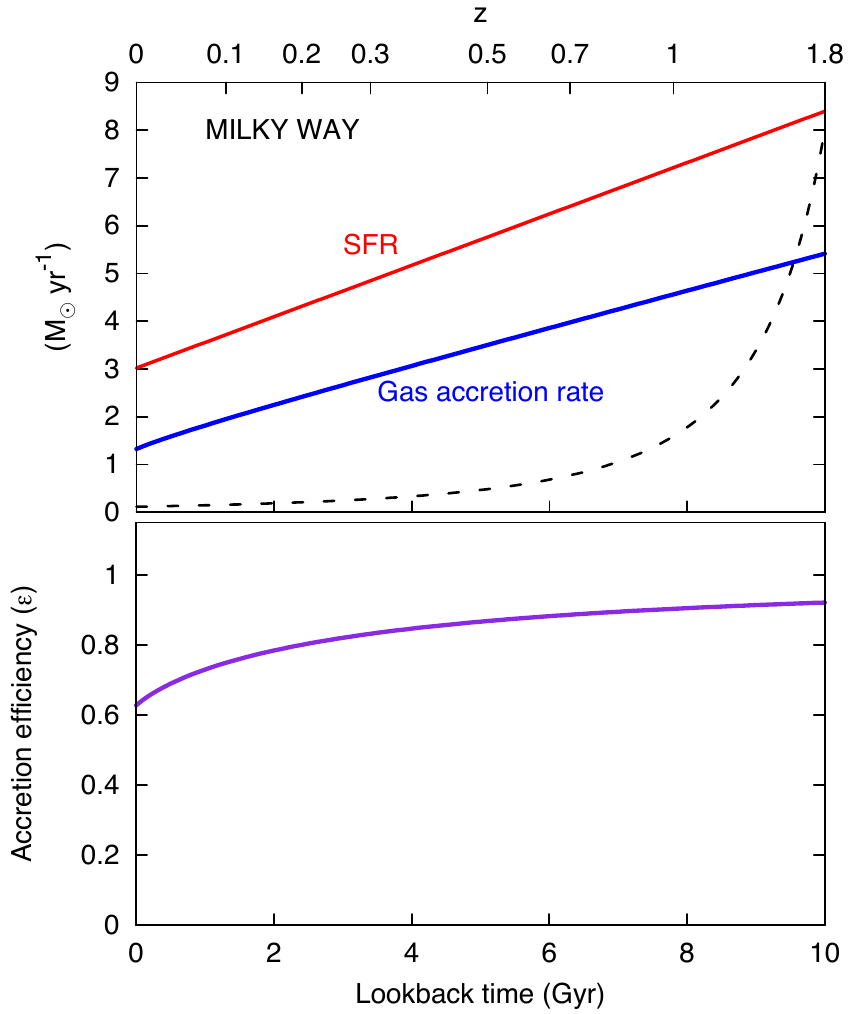}
\caption{
\emph{Top panel}: SFR and global gas accretion rate, versus time and
  redshift for the disc of the Milky Way. 
The dashed line shows the evolution of a closed box.
\emph{Bottom panel}: Gas accretion efficiency as a function of time,
$\varepsilon=1$ corresponds to a complete replenishment and thus a constant SFR.
\label{fig:sfh+accMW}}
\end{figure}

\begin{figure}
\centering
\includegraphics[width=0.5\textwidth]{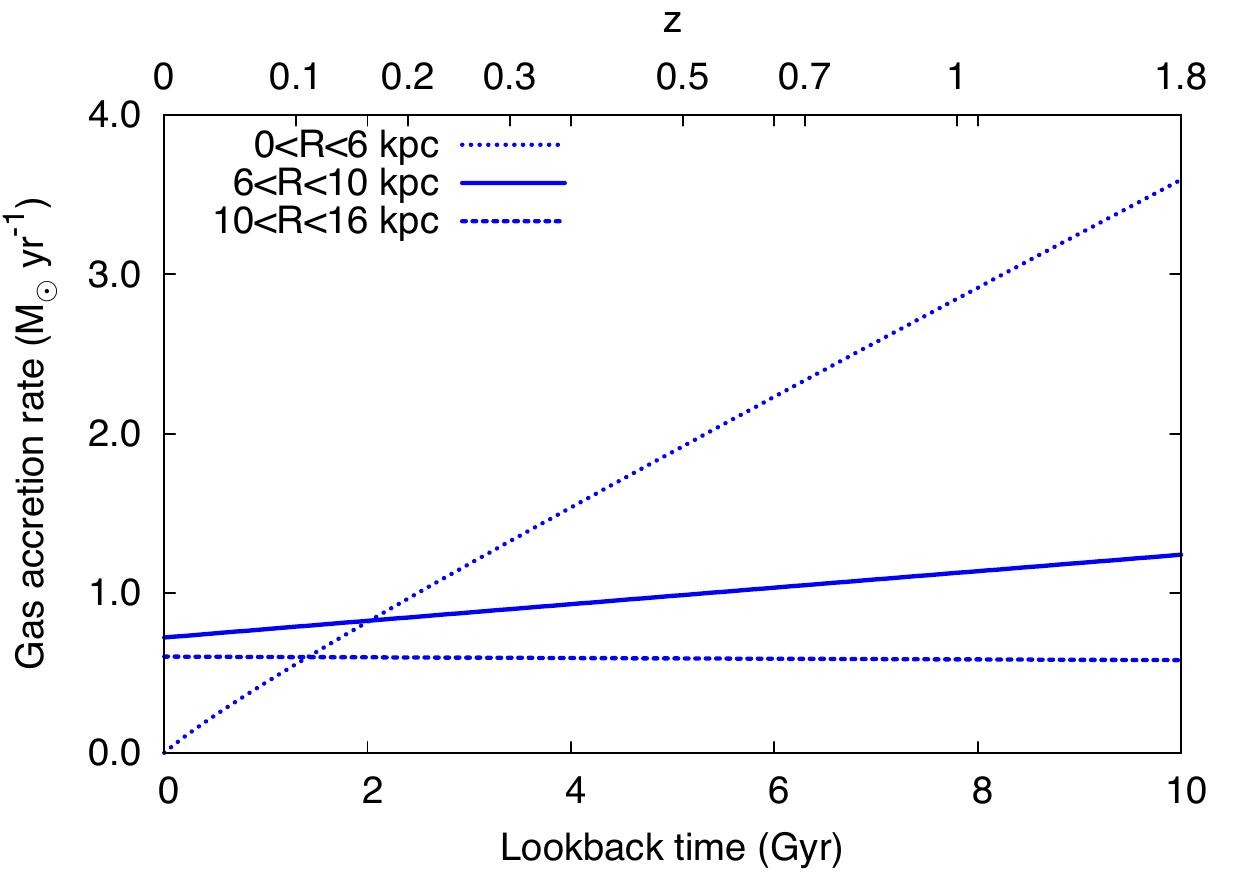}
\caption{
Gas accretion into the Milky Way's disc integrated over three annuli
  centered at $3$ kpc (\emph{short dashed line}), $8$ kpc (\emph{solid line})
  and at $13$ kpc (\emph{long dashed line}) versus time and redshift. 
\label{fig:3radii}}
\end{figure}

The SFR(t) of the Galaxy reconstructed using the linear temporal
dependency of eq.\ (\ref{eq:f1}) is shown in Fig.\ \ref{fig:sfh+accMW}
(top panel) ($\gamma = 2.8$).
We remind that the key parameters to obtain this curve are the current
SFR and the current stellar mass of the disc (eq.\ \ref{eq:gamma}). 
Changing these parameters and the formation time of the disc ($t_{\rm
  form}$) would produce rather intuitive variations in the value of
$\gamma$ and the steepness of the SFH. 
Extreme values can be obtained by taking ${\rm SFR}(0)=2 \moyr$,
$M_*(0)=5 \times 10^{10} \mo$ and $t_{\rm form}=8 \Gyr$ on the one
hand and ${\rm SFR}(0)=4 \moyr$, $M_*(0)=3 \times 10^{10} \mo$ and
$t_{\rm form}=12 \Gyr$ on the other, corresponding to $\gamma_{\rm
  max} = 7.9$ and $\gamma_{\rm min} = 0.8$ respectively.

Having derived $\gamma(R)$ we can proceed to
estimate the gas infall rate at each radius and its global value
by integrating over the whole disc surface using eq.\ \ref{eq:garf1_1}. 
Fig.\ \ref{fig:sfh+accMW} shows the global gas accretion rate
necessary to maintain the star formation in the Milky Way disc.
The accretion rate is positive at every redshift starting from
$5.4 \moyr$ at $t_{\rm form}$ and decreasing to
$1.3 \moyr$ at the current time.
It follows very closely the shape of the SFR.
These results are in agreement with earlier determinations based
on chemical evolution models \cite[e.g.][]{Tinsley80, Tosi88a}.
In the same figure, we also plot the SFR that our Galaxy
would have if it evolved like a closed system starting from the
same initial conditions.
A closed system is only able to produce $\simeq8.4\times10^9 \mo$ of stars
from $t=t_{\rm form}$ to $t=0$, virtually leaving no
remnant gas mass at present.
On the contrary, our model with gas infall predicts a present total gas mass of
$7.1\times10^9 \mo$, comparable to what observed.
The bottom panel of Fig.\ \ref{fig:sfh+accMW} shows the gas accretion efficiency
defined as the ratio of the accretion rate over the net gas-consumption including
feedback, $\varepsilon(t)=\dot M_{\rm ext}(t)/(1-\mathcal{R}){\rm SFR}(t)$.
For the whole disc of the Milky Way we obtain $\varepsilon_{{\rm MW}}(t)\sim0.6$ at $t=0$. 
$\varepsilon_{{\rm MW}}(t)$ has a mild dependence on time, reaching
a value of about 0.9 at $z=1$.
This shows that the efficiency of gas accretion is rather high but not
enough to keep the SFR constant ($\varepsilon=1$).
Thus, gas accretion does not feed the discs indefinitely, however it delays the
running-out of fuel significantly beyond the (closed-box) gas-consumption timescale.
The value of the current accretion rate and efficiencies are rather uncertain as they strongly depend on the assumed zero-points.
For instance, leaving all the other parameters unchanged and taking ${\rm SFR(0)}=2 \moyr$ would give a current accretion of only $0.14 \moyr$.
This large difference is totally localized to the current time, indeed the efficiency of accretion remains always above 0.5 from the disc formation to $z=0.1$ plunging then to 0.1 at $z=0$, see also the case of NGC\,5055 in Section \ref{sec:things}.

\begin{figure}
\centering
  \includegraphics[width=8.5cm]{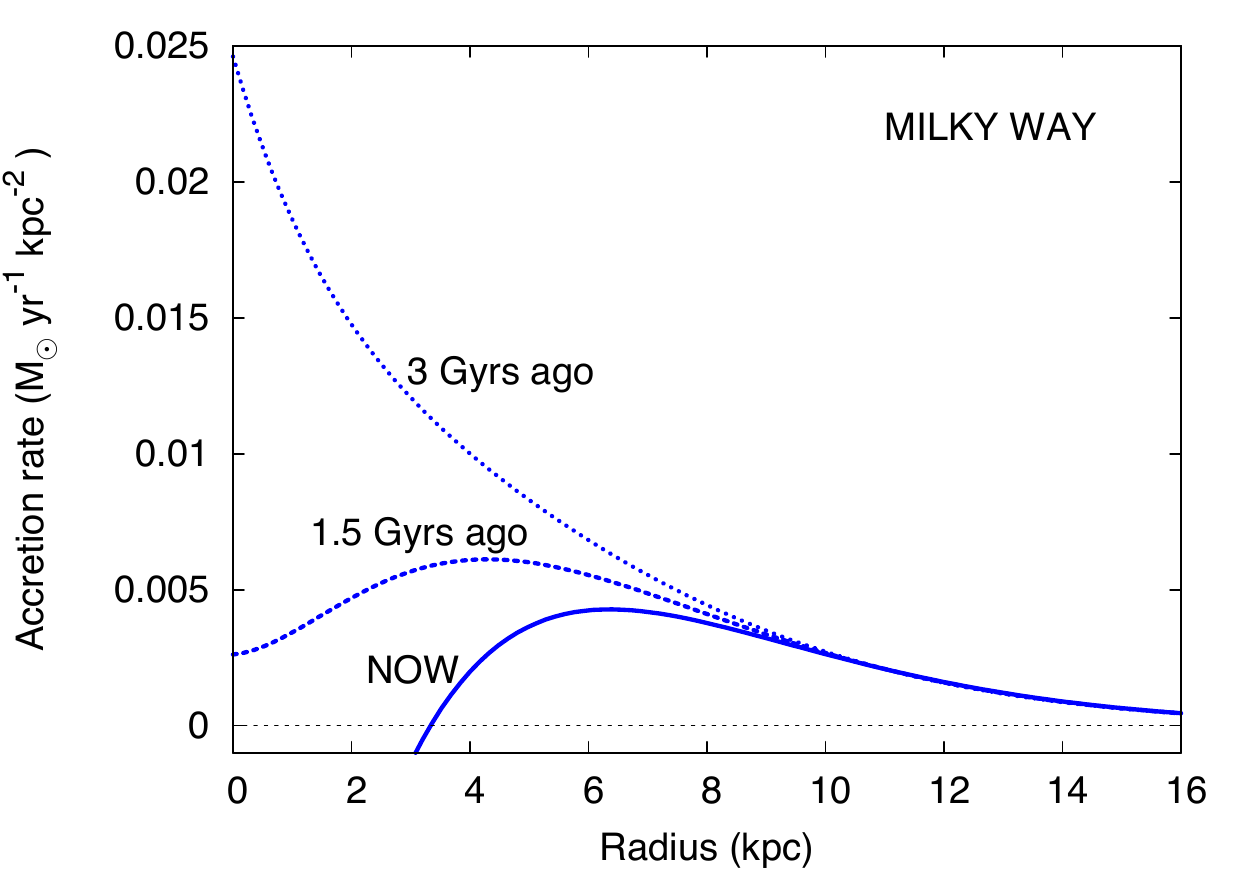}
  \caption{\label{fig:peak} 
Gas accretion profile at three epochs: lookback time = 3 Gyr, 1.5 Gyr, and now.
Note the very recent development of an inner depression and of a peak in the gas accretion distribution, now at about $R_{\rm peak}\simeq 6-7 \kpc$ from the Galactic centre.
}
\end{figure}

We now turn to the local infall rate $\dot \Sigma_{\rm ext}(R,t)$.
For visualization purposes, we split the galaxy in three
representative zones: the central part (between $0-6$ kpc), the solar
circle ($6-10$ kpc) and the outer disc ($10-16$ kpc).
For each region we calculate the local gas accretion rate integrating
over the annulus surface and we plot the results in
Fig.~\ref{fig:3radii}. 
We find three different regimes: i) in
the central region the gas accretion declines very steeply, 
ii) in the
solar circle the gas accretion declines by a factor 1.7, iii) in
the outer disc it increases slightly reaching its maximum at
$z=0$.
The behaviour visible in Fig.\ \ref{fig:3radii} reflects the simple
fact that the ratio between the stellar density and gas density
decreases strongly with radius in the Galaxy. 
The same kind of trends are visible in the SFH reconstructed for the
same annuli, given that as seen above, $\dot \Sigma_{\rm
  acc}(R,t)\sim(1-\mathcal{R}){\rm SFRD}(R,t)$. 

Fig.\ \ref{fig:peak} shows the gas accretion profile as a function of $R$ 
in the Milky Way's disc now and at two epochs in the recent past. 
The current profile is negative for $R<3 \kpc$, peaks at about $R = 6
\kpc$ and then falls further out.
The negative value in the central regions shows that there the amount of gas
is sufficient to sustain the star formation and the model would
predict outflow (of about $0.2 \moyr$).
This inner feature is of recent formation as 3 Gyrs ago the
peak of accretion was right in the centre.
So we conclude that the bulk of gas accretion is moving away from the inner
disc.
As a consequence star formation also moves out although with a delay with
respect to the accretion.
Note that the shape of the accretion profile at $t=1.5\,{\rm Gyr}$ is
remarkably similar to the current ${\rm SFRD}(R,0)$ (see Fig.\
\ref{fig:sfrdMW}), hinting at a delay between accretion and star
formation of a Gyr or so.
Interestingly, this is of the order of the gas depletion timescale.

We stress again that with our approach we are able to trace the
locations in the disc where the new gas begins to form stars. 
The infall itself could have happened somewhere else, realistically 
more further out and the infalling gas could have quiescently 
flown to inner radii before starting forming stars and becoming 
{\it detectable} by our method.

\subsection{External galaxies}
\label{sec:things}

\begin{figure}
\centering
  \includegraphics[width=8.5cm]{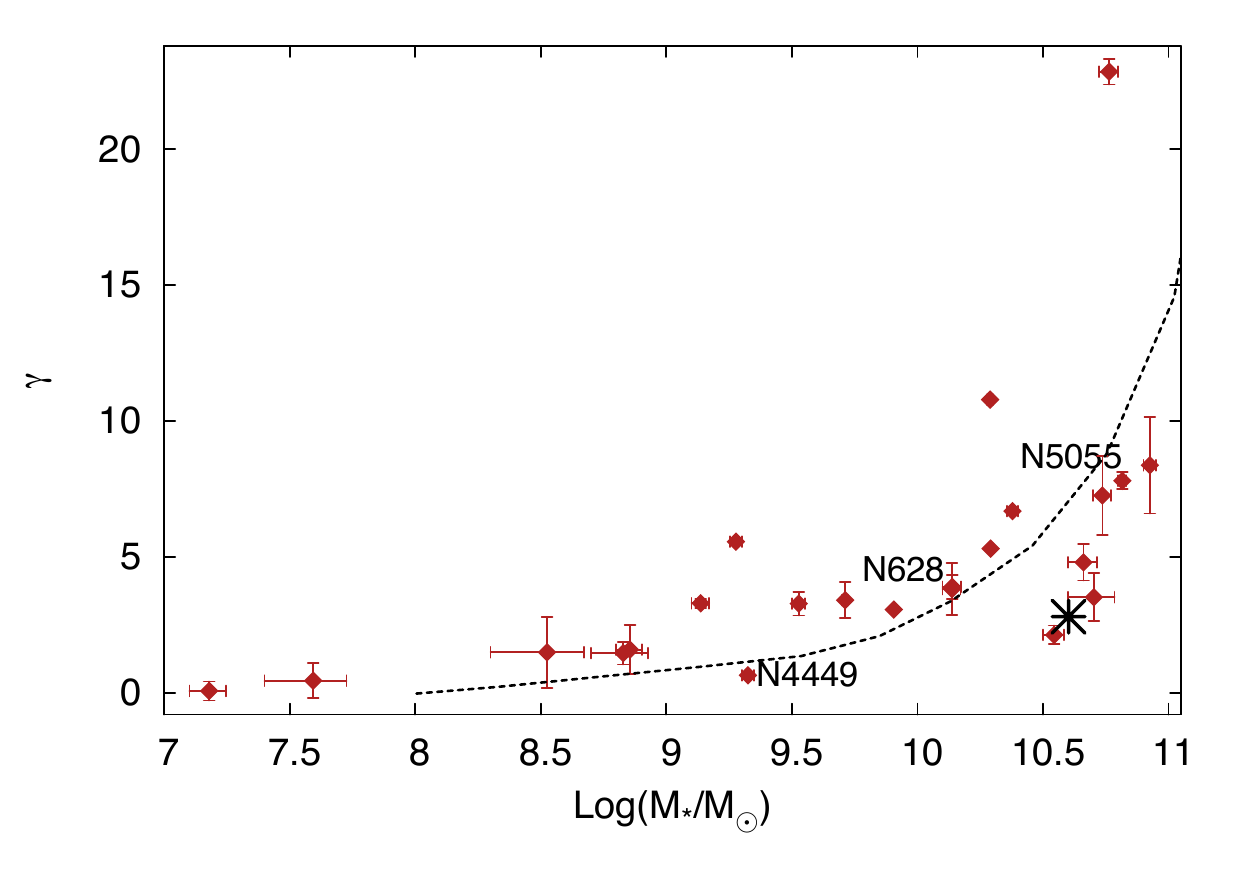}
  \includegraphics[width=8.5cm]{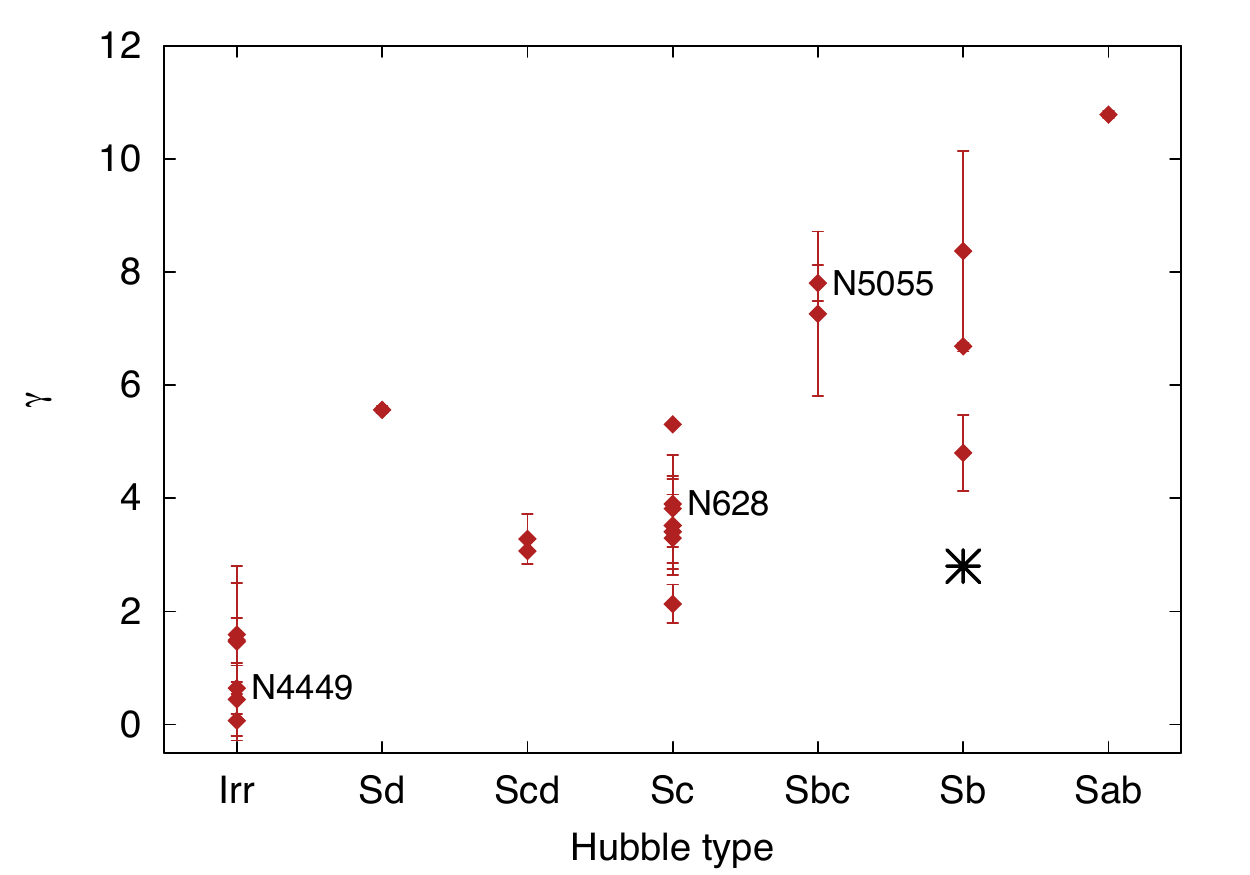}
  \caption{\label{fig:gammaLeroy} The (disc-integrated) parameter $\gamma$ (steepness of the SFH) 
for the sample of 21 disc galaxies of \citet{Leroy+08} as a function of the stellar mass (\emph{top panel}) and galaxy type (\emph{bottom panel}).
The star indicates the Milky Way.
A common time of disc formation of $10$ Gyr and the SFH-function $f(t)$ have been assumed for all galaxies.
The error-bars show the interval allowed by two different methods of integration described in the text.
The dashed curve in the top panel shows the value of $\gamma$ derived from the Scalo $b$-parameter for the SLOAN survey \citep{Brinchmann+04}.
}
\end{figure}

In this Section we extend the application of our method to 
a sample of external galaxies.
We consider galaxies from the THINGS survey \citep{Walter+08} 
for which radial profiles of gas density and SFRD are 
available \citep{Leroy+08}.
As for the Milky Way, we assume a common return factor
$\mathcal{R}=0.3$, an age of the discs of $t_{\rm form}=10$ Gyr and
the linear $f(t)$ described in Section \ref{sec:sfh}.
We derive the global values of $\gamma$ both using eq.\
(\ref{eq:gamma}) and by integrating $\gamma(R)$ (eq.\ \ref{eq:gammaR})
over the disc. 
In the former case the values of stellar masses and SFRs were taken from
Table 4 of \citet{Leroy+08}, in the latter we used their SFR
densities available online.
The two methods led, in some cases, to significantly
different values for $\gamma$.
This is partially due to the different range of integration,
\citet{Leroy+08} integrate out to $1.5\, R_{25}$ while we integrate
out to 5 scale-lengths (when the profiles extend this far, i.e. for
roughly for half of the galaxies) for consistency with the Milky Way
analysis.

\begin{figure*}
\centering
  \includegraphics[width=\textwidth]{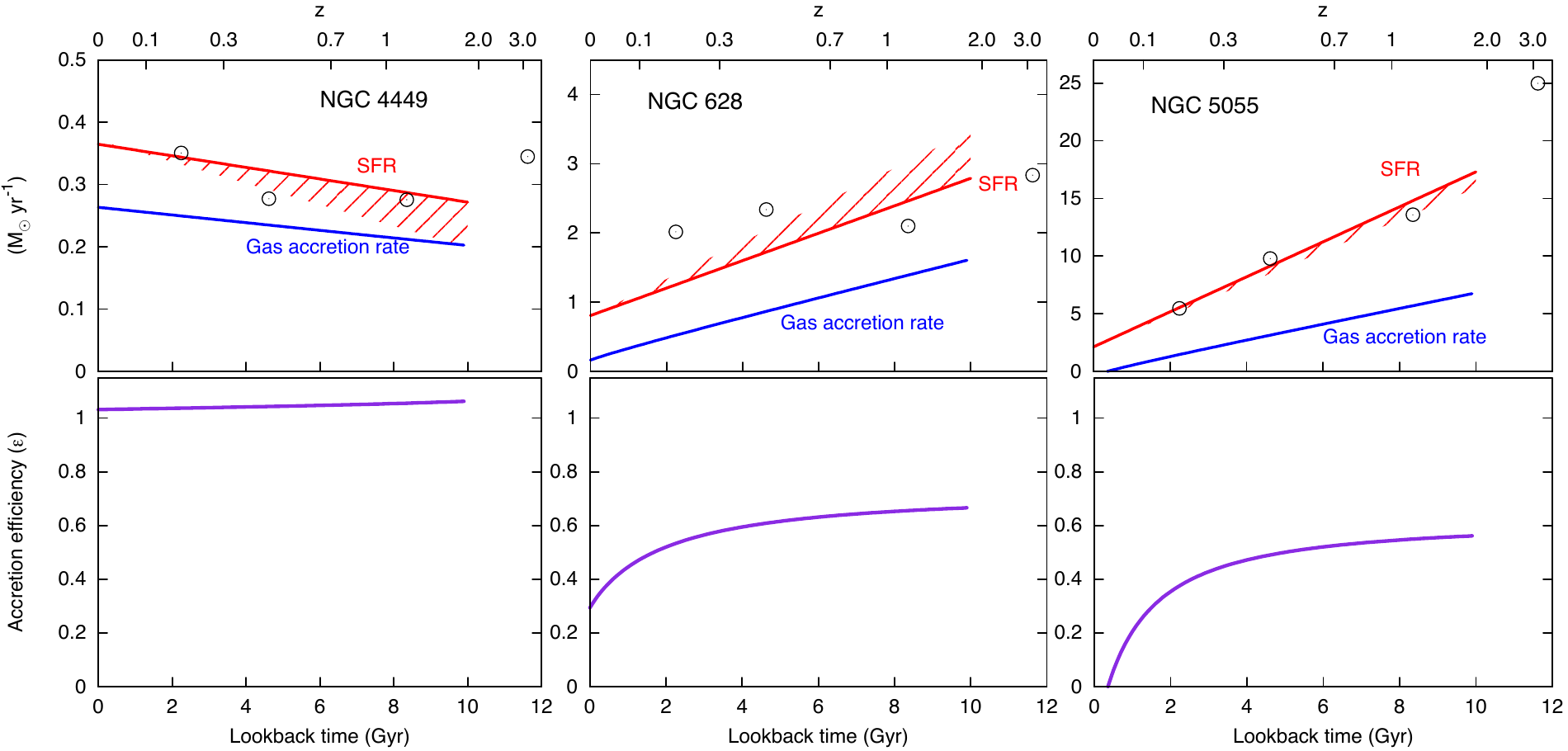}
  \caption{\label{fig:sfhsLeroy}
\emph{Top panels}: SFRs and gas accretion rates as a function of time for three representative galaxies from the sample of \citet{Leroy+08}.
The shaded areas show the uncertainties in the determination of the global $\gamma$ described in the text.
The open dots are the average SFHs estimated by \citet{Panter+07} for galaxies of the same stellar masses.
\emph{Bottom panels}: Gas accretion efficiencies as a function of time,
$\varepsilon=1$ corresponds to a complete replenishment and thus a constant SFR with time.
}
\end{figure*}

Fig.~\ref{fig:gammaLeroy} shows the derived values for the global $\gamma$ as
a function of the stellar mass (top) and Hubble type (bottom), the
black star shows the Milky Way.
The bars show the ranges allowed by the two methods described above. 
Note that for some galaxies also the stellar masses resulting from our integrations differ from those of \citet{Leroy+08}.
The values of $\gamma$ inferred for these galaxies are all
included between 0 and 11, with one outlier at $\gamma =23$, NGC\,2841, a massive Sb galaxy with rather low current SFR (shown only in the top plot).
There is a clear trend with stellar mass and Hubble type, showing that
big galaxies had much larger SFRs in the past. 
On the other hand, low values of $\gamma$, typical for smaller
galaxies betray a disc continuously forming through the acquisition of
fresh gas. 

As mentioned, we can relate our $\gamma$ to the Scalo $b$-parameter.
We use the recent determination of $b$ versus stellar mass
relation from the SDSS \citep{Brinchmann+04} and convert it into
$\gamma$.
We plot this relation as a dashed curve in the top panel of Fig.\
\ref{fig:gammaLeroy}.
The two parametrisations agree remarkably well, considering that
they are derived in completely different ways.
Our values tend to be slightly above the curve for masses lower 
than 10.5 in the log and below for larger masses.
This is likely due to the shape adopted for the SFH, see discussion in Section \ref{sec:fs}.
Note that, in this respect, Fig.~\ref{fig:gammaLeroy} is analogous to Fig.\ 3 in \citet{Kennicutt98b}.

For the above disc galaxies we have performed the same analysis as for the Milky Way.
We derived the radial profiles of $\gamma(R)$ and $\Sigma_{\rm ext}(R,t)$ and integrated the latter
to derive the global accretion rates.
Fig.\ \ref{fig:sfhsLeroy} shows the results for three galaxies representative of three classes
of stellar masses and galaxy types (see labels in Fig.\ \ref{fig:gammaLeroy}).
In general, small late-type galaxies tend to have a flatter SFH and an
efficiency of gas accretion close to 1, see the case of NGC\,4449
(left panels). 
Galaxies of intermediate types have a shallower SFR and efficiency
gradually decreasing with time.
Finally, massive discs with low current star formation rates seem to
have exhausted their capability of acquiring significant amount of gas
from the environment. 
They now appear to be consuming their remnant gas before moving to the red sequence \citep[e.g.][]{Cappellari+11}.
In the same Fig.\ \ref{fig:sfhsLeroy} we also compare our results with
the average SFHs derived using the SLOAN survey by \citet{Panter+07},
shown as open circles.
Panter et al.\ divide their galaxies in bins of current stellar
masses, those relevant here are $3\times10^{10}<M_{*}<1\times10^{11}$,
$1\times10^{10}<M_{*}<3\times10^{10}$, $M_{*}<1\times10^{10}$,
respectively applicable for NGC\,5055, NGC\,628, and NGC\,4449. 
The agreement between our SFRs and their average SFRs is very good,
except for the galaxies in the intermediate mass class as our method
tends to find steeper SFHs. 
Note however that this may also depend on the way the stellar masses are
estimated.
We used the estimate of \citet{Leroy+08} who simply multiply the
3.6-$\micron$ surface brightness by a constant mass-to-light ratio of
0.5. 

In general, the analysis of these 21 THINGS galaxies gives results
similar to those found for the Milky Way's disc. 
Gas accretion proceeds almost parallel to the SFR and it remains
significantly important throughout a large fraction of the 
lifetime of a typical disc galaxy. 
The efficiency of accretion seems to decrease fast for large
galaxies, while small late-type discs remain quite efficient in
collecting cold star-forming gas from the environment.
The current ($z=0$) values of gas infall that we derive are quite
low for most galaxies in the sample and a few also have negative values
showing that the available gas is more than sufficient for star
formation to proceed.
Thus, it would appear that these disc galaxies do not need
gas accretion to sustain their current star formation (see e.g.\ NGC\,5055).
This is potentially a very interesting result but unfortunately,
as shown for the Milky Way, these values are very uncertain because they come from an 
extrapolation.
Moreover, they depend on parameters like the stellar masses 
that have intrinsic uncertainties and other parameters that are
simply assumed (see eq.\ \ref{eq:garf1_1}).
For instance if one increases the formation time of the disc or
decrease the return factor the value of the current accretion rate may
increase substantially.
What remains indisputable is the fundamental relation between star
formation and accretion and the fact that the latter $must$ be a major
player in the evolution of galaxies from their formation to the
present. 
There is no way a disc galaxy could become as we see it today without
substantial gas infall at $z<1$.

\section{Discussion}\label{sec:discussion}

In the previous Sections we have shown how a simple comparison between the stellar density in galactic discs and their current SFRD (or gas density) implies the need for a large amount of gas infall during their lifetimes.
This need is rooted in the existence of the Kennicutt-Schmidt law and in the assumption that this does not significantly evolve with time.
Under this assumption we were able to derive the accretion rates as a function of galactocentric radius and time for a number of disc galaxies including the Milky Way.
Here we discuss the limitations of our approach and the implications of our results.

\subsection{Delayed stellar feedback}\label{sec:nonIRA}

Treating stellar feedback with the I.R.A.\ saves computing time and it allows us to write simple analytic expressions for the equations of the model, however the delayed return from stars with mass $M<8 \mo$ may be an important effect that needs to be quantified.
We added this ingredient using the approximation that a star with mass $M$ returns a fraction $\eta(M)$ of its mass to the interstellar medium at a time $t_{\rm MS}(M)$ \citep[see][]{Kennicutt+94}.
The stellar lifetime ($t_{\rm MS}$) are taken from \citet{Maeder&Meynet89}:
\begin{eqnarray}\label{eq:tMS}
  &t_{\rm MS} (M)~~= &12\, \left(\frac{M}{\rm M_\odot}^{-2.78}\right)\,\Gyr ~~~~M\leqslant10\mo \nonumber\\
  & &0.11 \, \left(\frac{M}{\rm M_\odot}^{-0.75}\right)\,\Gyr ~~~M>10 \mo
\end{eqnarray}
given that $t_{\rm MS} (1$ M$_\odot)>10$ Gyr stars below $\sim 1$ M$_\odot$ do not contribute to the feedback.
The difference between the I.R.A.\ and the delayed return is that the term $\dot\Sigma_{\rm fb}$ in eqs.\ (\ref{eq:tinsley1}) and (\ref{eq:tinsley2}) becomes:
\begin{equation}\label{eq:sigmaFB_DF} 
\dot\Sigma_{\rm fb}=\int_{M_{\rm min}(t_{\rm form}-t)}^{100}{{\rm SFRD} (t+t_{\rm MS}(M))}\eta(M)\phi(M){\rm dM}
\end{equation}
where $\phi(M)$ is the IMF.
The integration is performed from $M_{\rm min}(t_{\rm form}-t)$, which is the mass of a star with a MS time equal to the age of the disc at that time \citep{Prantzos08}; note that with $t$ we indicate, as usual, lookback time.
In order to calculate $\gamma(R)$ with delayed feedback, we have numerically intergrated eq.\ (\ref{eq:tinsley1}) from the present to $t_{\rm form}$ using eqs.\ (\ref{eq:sfrd}) and (\ref{eq:f1}) for ${\rm SFRD}(R,t)$.
At each radius $R$, the integration has been repeated iteratively until we matched the observed stellar surface brightness $\Sigma_*(R,0)$.
Then the global $\gamma$ was calculated using eq.\ (\ref{eq:gamma}).
Finally the accretion rate as a function of radius and time was derived using eq.\ (\ref{eq:garf1_1}). 

\begin{figure}
\centering
  \includegraphics[width=8.5cm]{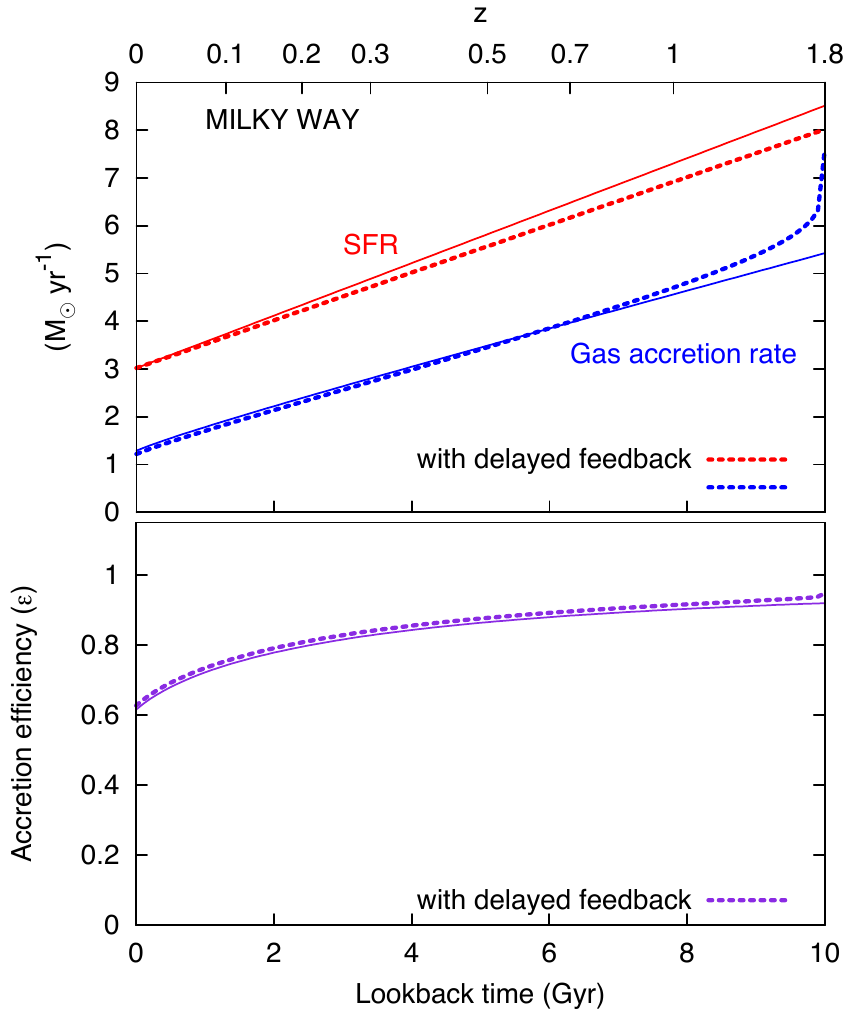}
  \caption{\label{fig:nonIRA}
\emph{Top panel:} SFR and global gas accretion rate, versus time and
  redshift for the disc of the Milky Way with instantaneous recycling approximation (thin solid lines) and delayed feedback (thick dashed lines).
\emph{Bottom panel}: Gas accretion efficiency as a function of time with instantaneous recycling approximation (thin solid line) and delayed feedback (thick dashed line).
}
\end{figure}

Fig.\ \ref{fig:nonIRA} shows the comparison between the global SFR and accretion rate for the Milky Way with I.R.A.\ and with delayed stellar feedback.
The values of current SFR and stellar mass are kept fixed in both cases.
As expected the global gamma with delayed feedback is lower, $\gamma_{\rm DF}=2.7$ (2.6 for a Chabrier IMF), than that with I.R.A.
This is because if the mass return to the ISM is delayed then fewer stars need to form in order to have the same stellar mass at the present time.
The gas accretion is higher at the beginning due to the fact that less gas is available from stellar recycling and becomes slightly lower later on when feedback from low mass stars becomes important.
The bottom panel of Fig.\ \ref{fig:nonIRA} shows the accretion efficiency calculated with and without delayed feedback.
Accretion with delayed feedback tends to have a slighly larger efficiency.
However the shape of the two curves is exactly the same and the difference in the normalization is minor.
In conclusion, including delayed feedback in our calculations makes little difference and it goes in the direction of increasing the efficiency of gas accretion.

\subsection{Parametrisation of the SFH}\label{sec:fs}

The reconstruction of the star formation rate density as a function of time requires the parametrisation of the SFH.
In Section \ref{sec:sfh} we have chosen a simple first order polynomial.
We also point out that a more general power-law would give similar dependencies for $\gamma(R)$.
Here we consider three different parametrisations and show that our results do not change significantly.
We take the following functional forms:
i) a piecewise-linear function of time $f_2(t)$ that linearly increases to a maximum SFR at $t=t_{\rm c}$ and remains constant at higher lookback times;
ii) a triangular function of time $f_3(t)$ that linearly increases to a maximum SFR at $t=t_{\rm c}$ and then decreases to ${\rm SFR}(0)$ at $t=t_{\rm form}$;
iii) an exponential function of this form:
$ f_4(t) = e^{t/\tau}$
where $\tau$ is the exponential $e$-folding time of the SFR.
With these choices we encompass both monotonic and non-monotonic functions.
In ii) $\gamma$ is defined as the ratio between ${\rm SFR}(t_{\rm c})/{\rm SFR}(0)$. 

In Table \ref{tab:fts}, we show the values of the global $\gamma$, and the present accretion rate $\dot M_{\rm ext}(0)$ and accretion efficiency $\varepsilon(0)$ obtained with the four types of $f(t)$ for the Milky Way.
We took $t_{\rm c}=7.7 \Gyr$ ($z_{\rm c}=1$).
For the exponential function, often used for early type galaxies \citep[e.g.][]{Bell+03},
we found a rather long scale-time of $\tau= 8.5\Gyr$ and a shape of the SFH very similar to that obtained with $f_1(t)$. 
Even though these three functions depict very different trends for the SFH, the disc-integrated $\gamma$ varies only by $\sim 17$\% and
the values of the present accretion rate and $\varepsilon(0)$ change by no more than 13 \%. 
Note that the inverse of the Scalo $b$-parameter would correspond to a piece-wise linear functional form $f_2(t)$ for $t_{\rm c} \to 0$.
All the above values are obtained using the I.R.A.
Using the delayed return with different shapes of the SFH can produce different effects.
For instance an exponential SFH of the type $f_4(t)$ maximizes the gas consumption timescales, which can increase up to a factor $2-3$ at the earliest times \citep[see][]{Kennicutt+94}.

\begin{table}
\centering
\begin{tabular}{|l|l|l|l}
\hline
$f(t)$ & $\gamma$ & $\dot M_{\rm ext}(0)$ & $\varepsilon(0)$ \\
\hline
linear (1) & 2.8 & 1.31 & 0.62 \\
piece-wise linear (2) & 2.5 & 1.27 & 0.61\\
triangular (3) & 2.4 & 1.21 & 0.58\\
exponential (4) & 3.2 & 1.49 & 0.71 \\
\hline
\end{tabular} \newline
\caption{ \label{tab:fts}
The effect of using 
different parametrisations for the SFH described in the text. $\gamma$
is always defined as the ratio between the initial over the current
SFR except for the triangular function where it is the ratio between
the SFR at $z_{\rm c}=1$ over the current value.
}
\end{table}

\subsection{Effect of stellar migration}\label{sec:migration}

In the above we have assumed that radial motions of stars in the disc are not significant, here we relax this assumption and estimate the effect of a global migration \citep[e.g.][]{Roskar+08,Minchev+11}.
For simplicity we take an exponential stellar profile with central surface brightness $\Sigma_{*}(0)$, a scale length $h_R$, and a maximum extent at radius $R_{\rm m}=5\,h_R$.
The stellar mass of the disc is:
\begin{equation}\label{eq:massexp}
 M(\Sigma_*(0),h_R)=2\pi\Sigma_{*}(0)h_R^2\left[1-e^{-\frac{R_{\rm m}}{h_R}} \left(1+\frac{R_{\rm m}}{h_R} \right) \right]
\end{equation}
Consider now a radial motion with a constant velocity $v$ at $R_{\rm m}$ that lasts a time $t$ directed inward or outward. 
Since the initial stellar profile is truncated beyond $R_{\rm m}$, all stars at this radius will move to the new position $R_{\rm m}+vt$, which represents a new maximum radius and is greater or smaller than $R_{\rm m}$ depending on the sign of $v$. 
This condition translates to the following relation involving the new (i.e.\ after migration) stellar profile $\Sigma_*^{\prime}$:
\begin{equation}\label{eq:sigma1}
 \Sigma_*^{\prime}(R_{\rm m}+vt)=\Sigma_*(R_{\rm m})
\end{equation}
that, together with the assumption that the profile remains exponential and that the mass is conserved:
\begin{equation}\label{eq:mass1}
 M^\prime(\Sigma_*^\prime(0),h_R^\prime)=M(\Sigma_*(0),h_R),
\end{equation}
allows us to estimate $\Sigma_*^{\prime}(0)$ and $h_R^\prime$, the new central surface brightness and scale length.

In Table \ref{tab:rad} we report the variation of the exponential profile of the Milky Way ($\Sigma_*(0)=648.2 \mopc$ and $h_R=3.2$ kpc) in the presence of coherent radial motions. 
When the flow of stars is directed inward, obviously the normalization of the stellar profile increases and the scale length decreases.
Eq.~(\ref{eq:gammaR}) tells us that $\gamma(R)+1\propto \Sigma_*(R)$. Therefore, the profile of $\gamma(R)$ will change to $\gamma^\prime(R)$ according to this relation:
\begin{equation}\label{gammarad}
 \gamma^\prime(R)=\frac{\Sigma_*^\prime(R)}{\Sigma_*(R)}(\gamma(R)+1)-1
\end{equation}
For radii $R$ where $\Sigma_*^\prime(R)/\Sigma_*(R)>1$, eq. (\ref{gammarad}) predicts an increase in the parameter $\gamma(R)$ and vice versa for $\Sigma_*^\prime(R)/\Sigma_*(R)<1$.
This can be explained considering that a radial motion that brings more stars at a certain radius (i.e. $\Sigma_*(R)^\prime/\Sigma_*(R)>1$) makes the latter as it had been {\it more star-forming}.
Despite the changes in $\gamma(R)$ it is remarkable that the new global $\gamma$ changes very little, see the rightmost column of Table \ref{tab:rad}, showing that our global results are only slightly affected by radial migration.

\begin{table}
\centering
\begin{tabular}{|l|l|l|l|}
\hline
 $vt{\rm[kpc]}$&$\Sigma_*^\prime$ & $h_R^\prime$ & $(\gamma^\prime/\gamma)_{\rm MW}$\\
\hline
\hline
-3 & 1226.96 & 2.31 & 1.25\\
-2 & 980.63 & 2.59 & 1.08\\
-1 & 792.97 & 2.88 & 1.02 \\
0 & 648.22 & 3.2 & 1.00\\
1 & 534.77 & 3.54 & 1.04\\
2 & 444.79 & 3.89 & 1.12\\
3 & 372.61 & 4.27 & 1.21 \\
\hline
\end{tabular} 
\caption{\label{tab:rad}The effect of radial migration of stars on an exponential stellar profile and on the disc-integrated $\gamma$ for our Galaxy}
\end{table}

\citet{Sellwood&Binney02} showed that the dominant effect of spiral waves in galaxy discs is to churn the stars in a manner that preserves the overall angular momentum distribution. 
Stars move both inwards and outwards without any significant radial spreading of the disc or increase in non-circular motions.
In their Fig.\ 13 they plotted the final distribution of home radii for six different radial bins for our Galaxy. 
The distributions are roughly Gaussian for every radial bins, with the largest dispersion
$\sigma\simeq3 \kpc$ at about $R_\odot$. 
So as to give an upper limit on this effect we infer the new stellar profile $\Sigma_*(R)^\prime$ as the average of that modified by inward radial motion with parameter $vt=-3$ kpc and that of an outward motion with parameter $vt=3$ kpc.
We found that the changes in $\gamma(R)$ are significant for some radii but there is basically no effect on the global value of $\gamma$.
We conclude that both migration and random redistribution of stars over the disc do not affect our global results.

\subsection{Comparison to other studies}
\label{sec:Others}

It is a matter of debate whether galaxies had their reservoir of gas in place from the beginning or they have been harvesting the gas they needed throughout their lives.
We find that the second must be the case for all late type galaxies.
A concentration of gas as large as the current disc stellar masses would imply an extremely high SFR at earlier times, a sudden formation of most of the stars and a steeply declining star formation at later times.
This is how a red sequence galaxy forms and evolves.
Galaxies like the Milky Way or spirals of later types clearly did not have this kind of star formation history.
The evidence comes from the reconstruction of the star formation histories from the color magnitude diagrams \citep[e.g.][]{Harris+09}, from chemical evolution models \citep[e.g.][]{Chiappini+01}, and from cosmology \citep[e.g.][]{Hopkins+08}.
Our study further supports this point making use of the Kennicutt-Schmidt law.

A possible way out would be to invoke a variation of the K-S law with redshift or with metallicity as proposed by e.g.\ \citet{Krumholz&Dekel11}.
However, to reconcile the observations with negligible accretion at $z<1$ the variations in the K-S law should be very large.
If one required the initial gas mass of the Milky Way disc to be as large as the stellar mass today, the K-S law would imply an initial SFR of $\sim50 \moyr$.
At that rate, the stellar disc would form in a Gyr and the subsequent evolution would be passive (similar shape to the closed-box shown in Fig.\ \ref{fig:3radii}).
Therefore to keep the current SFR at the observed value the transformation of gas into stars should have been an order of magnitude less efficient than the K-S law would predict.
Moreover, this efficiency should have evolved with the galaxy itself in order to keep the star formation roughly constant in time.
This is in contradiction with the fact that today galaxies with very different masses and thus different disc metallicities roughly follow the same K-S law \citep{Leroy+08}.
Thus, it seems unlikely that any realistic evolution of the K-S law can account for the evolution of discs without conspicuous gas accretion at $z<1$.

Despite the lack of metallicity information in our analysis, we have tried to quantify the above effect by assuming a different star formation law for low density environments.
To this aim, we broke the star formation law into two parts: we assumed that the standard K-S is valid only down to a break surface density of 10 $\mopc$ in hydrogen, below which the law takes a different form.
We took this form from the work of \citet{Roychowdhury+09} who studied a compilation of extremely faint (and presumably metal poor) dwarf galaxies and found that a normalization of $A=1.0\times10^{-5}$ and a slope of $N=2.47$ are suitable for these low density environments.
This parametrisation makes star formation less efficient at low gas densities.
We repeated all the above calculations and found that the general result is, as expected, that more accretion is required, because star formation needs more gas to occur at the same rate.
However, the differences are not big.
For instance the amount of gas in the Milky-Way disc at all times increases by a factor in the range $9-18\, \%$ and the amount of accretion today only by $5\,\%$.
For the other galaxies the differences can, in some cases, go up to a factor $\sim 2$.
The reason for this is that the second term on the r.h.s.\ of eq.\ (\ref{eq:garf1_1}) does not change by more than a factor 2.3 below the break surface density.

Another important result of our investigation is that gas accretion moves in time from the central to the outer parts of the discs.
In other words, the inner parts dominate the accretion at earlier times, whilst the outer parts dominate at later times, see Fig.\ \ref{fig:3radii}.
This inside-out evolution is also responsible for the production of metallicity gradients \citep[see e.g.][]{Matteucci+99,Boissier&Prantzos99}.
The progressive shift of accretion may be related to the angular momentum of the infalling material and it would be interesting to check it against the results of ab-initio hydrodynamical simulations \citep[e.g.][and references therein]{Piontek&Steinmetz11}.

\begin{figure}
\centering
  \includegraphics[width=8.5cm]{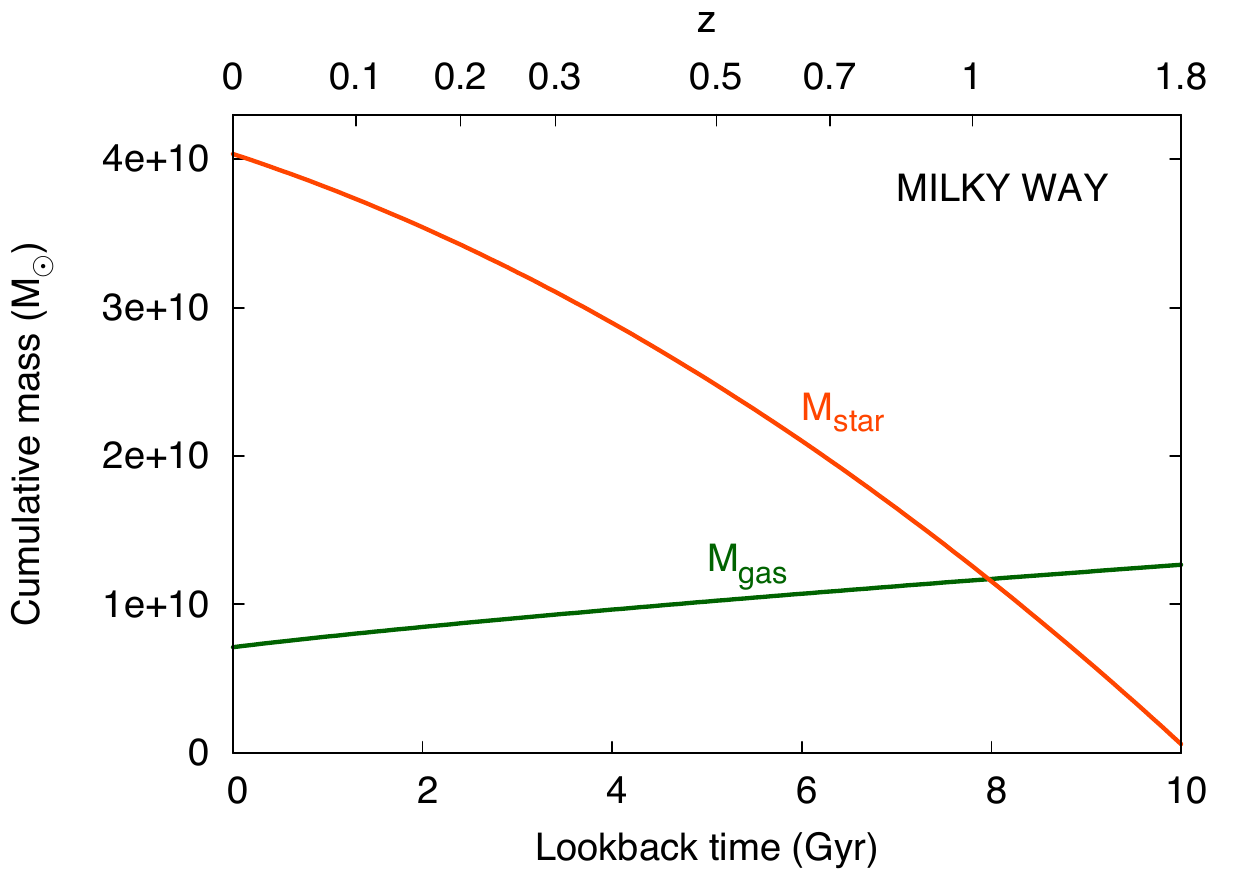}
  \caption{\label{fig:massEvolution} Gaseous and stellar mass as a function of the lookback time in the disc of the Milky Way.}
\end{figure}

Our model also predicts the distributions of stars and gas in the disc at all times.
We found that the stellar disc of the Milky Way remains roughly exponential, at least in the inner parts, with a scale-length slightly increasing with time from 2.7 to $3.2 \kpc$.
The gaseous disc, initially exponential, develops an inner depression with time, as stars are formed in the central region and the accretion becomes more and more efficient in the outer parts.
Fig.\ \ref{fig:massEvolution} shows the evolution of the stellar and gaseous masses in the Milky Way disc.
The cumulative stellar mass increases slowly with time.
Roughly $2/3$ of the stars formed at $z<1$ as a consequence of gas accretion towards the inner disc.
On the other hand, the gas mass remained almost constant and always rather low.
Interestingly, the ratio between our initial and current gas mass for the Milky Way is the same as that obtained by comparing the \hi\ mass in galaxies at $z=0$ with that of Damped Ly$\alpha$ systems \citep[e.g.][]{Zwaan+05, Lah+07}.
We stress again that the gas total mass in Fig.\ \ref{fig:massEvolution} is calculated within $R_{\rm m}=5\,h_{\rm R}$, more gas may lay at larger radii and flow inwards with time.

In a recent paper \citet{Naab&Ostriker06} built a model for the evolution of the Milky Way that has aspects in common to our investigation.
They consider the growth of a disc that goes through an initial merger phase that lasts about $2\Gyr$ and then follows a quiescent evolution.
This second phase is analogous to the phase we are considering in this paper.
\citet{Naab&Ostriker06} assume the variation of the scalelength of the disc with time as the inverse of the Hubble parameter, so by a factor about 3 in the last $10 \Gyr$.
This assumption together with the K-S law, used to derive the SFR, determine the amount of gas accretion onto the disc.
In contrast to this approach, we derive the ${\rm SFRD}(R,t)$ from the normalization at redshift zero and use the K-S law to derive the amount of gas available at any time and the amount of gas needed (gas accretion) to reproduced that ${\rm SFRD}(R,t)$.
Then from this we obtain the evolution of the stellar and gaseous discs.

From the above it is clear that the methods used in the present paper and in \citet{Naab&Ostriker06} are rather complementary and it is remarkable that we attain a very similar evolutionary picture for the Galaxy disc.
The main difference is the evolution of the stellar-disc scalelength, which evolves qualitatively in the same way but quantitatively much less in our treatment, i.e.\ our disc has initially $h_R \simeq 2.7 \kpc$ as opposed to $h_R \sim 1 \kpc$ in their model (see their figure 1).
The SFRs in the Solar neighbourhood are slightly different but both comparable with the determination of \citet{Rocha-Pinto+00}.
Finally, they also find that the accretion rate closely follows the SFR at every time and at present is about $2 \moyr$, slightly higher than ours.
Our analysis also shows similarities to the work of \citet{Boissier&Prantzos99}.
These authors choose a different form of the star formation law and derive chemical and spectrophotometric properties of the Milky-Way disc.
In their case, the accretion-rate is fixed and assumed to take an exponential law.
Globally, all these descriptions give a consistent picture of the evolution of a galactic disc, reinforcing the validity of the different parametrizations.

\subsection{What type of accretion?}

Our treatment does not include metallicity information but it is clear that
the infalling gas is bound to be at low metallicity.
The need for metal-deficient accretion has been shown and confirmed throughout
the last decades starting from pioneering works such as \citet{Tinsley&Larson78} and \citet{Tinsley80}.
More recently it has become clear that one needs two
epochs of gas infall: one at the early epochs and one slower and still ongoing, which is the one we are describing here \citep{Chiappini+97}.
The metallicity of this gas should be fairly low, of order $0.1\,Z_{\rm \odot}$ or less \citep[e.g.][]{Tosi88b}.
The origin of this accretion is still matter of debate and it is beyond the scope of this paper.
As mentioned, several numerical investigations show that cooling of the hot corona may provide a reasonable amount of gas accretion onto the Milky-Way disc, although at a relative low rate \citep{Peek+08, Kaufmann+09}.
These simulations roughly reproduce also the kinematics of this gas \citep{Kaufmann+06} and the radial distribution of the accretion \citep{Peek09}.
However, the whole mechanism of direct cooling via thermal instability has been recently challenged both by analytical analyses \citep{Binney+09, Nipoti10} and by recent grid-based simulations \citep{Joung+11} and the matter is still not settled.

Our accretion is smooth but this is obviously an artifact of our assumptions.
In principle the main contribution to gas infall could be minor mergers with gas rich satellites, which would give presumably a more bursty SFH.
In this sense, our analysis is still applicable if one averages all quantities over a time larger than the typical starburst timescale.
Current \hi\ observations seem to show that the amount of gas that can be realistically brought into large galaxies by gaseous satellites in the local Universe is rather low \citep{Sancisi+08}.
However, these estimates are still uncertain and only future \hi\ surveys will allow us to determine the accretion from mergers with the required accuracy.
A recent analysis of the contribution of satellites to the gas accretion in the Milky Way shows that they would also not reproduce the distribution of SFR observed today \citep{Peek09}.

Finally, we found that accretion and star formation proceed together.
This is clear from eq.\ \ref{eq:gar} and from Figs.\ \ref{fig:sfh+accMW} and \ref{fig:sfhsLeroy}.
This finding may just highlight the obvious point that the more a galaxy accretes gas the more stars it forms.
However, the regularity of this link seems to point at a more fundamental interplay between star formation and accretion \citep[see also the discussion in][]{Hopkins+08}.
In a series of papers \citep{Fraternali&Binney08, Marinacci+10, Marinacci+11} it has been suggested that gas accretion in discs is fundamentally related to star formation because it is produced by supernova ({\it positive}) feedback.
Supernovae eject (high-metallicity) gas clouds out of galactic discs mixing it with the surrounding hot corona and causing prompt cooling of the latter.
This implies an accretion that follows closely the star formation in the disc and proceeds from inside out given the distribution of cloud orbits \citep{Fraternali&Binney06}.
\citet{Marasco+11} applied this model to the Milky Way and found a global accretion rate of $\sim 2 \moyr$.
They also found that the current accretion profile is an increasing function of the radius out to a peak at $R\sim9 \kpc$ followed by a sudden drop further out.
In this paper, we found a similar shape but with a peak located closer in, see Fig.\ \ref{fig:peak}.
However, as mentioned, our approach traces the location where the new gas forms stars.
So it is conceivable that the infall took place at slightly larger radii, and then it flowed quiescently in.
It is compelling that two completely independent methods as these produce such similar accretion patterns and global gas accretion rates.

\section{Conclusions}\label{sec:conclusions}

In this paper we have proposed a simple method to derive the amount of gas needed for the star formation to proceed in a galactic disc, using the Kennicutt-Schmidt law.
We found that in typical disc galaxies (Sb or later types) most star-forming gas is not in place in the disc at the time of formation but needs to be slowly acquired from the surrounding environment.
We derived the gas accretion rate as a function of time for the Milky Way and other 21 disc galaxies from the THINGS sample.
We parametrized the SFH with a simple polynomial function with a parameter ($\gamma$) that defines the slope of the SFH, $\gamma=1$ stands for a flat SFH.
We summarize our results as follows:
\begin{enumerate}
\item the disc of the Milky Way as a whole (not only the Solar neighborhood) formed stars in the past a rate that was $\sim2-3$ times larger than now;
\item the steepness of the SFH is a function of galaxy mass and Hubble type, with late type galaxies having nearly flat or inverted SFH;
\item gas accretion is tightly linked to star formation, a constant ratio between the two is maintained for a large fraction of the life of a galaxy, pointing at a reciprocal interplay;
\item galaxy discs must have experienced accretion of large fraction of their mass (in gaseous form) after $z=1$, unless the K-S law had a strong evolution with $z$;
\item gas accretion progressively moves from the inner to the outer regions of galaxy discs, as a consequence star formation also proceeds inside-out;
\item the efficiency of gas accretion is less than 1 and decreases with time in all galaxies expect some late-type dwarfs, as a consequence the gas consumed by star formation is not completely replenished and the galaxy eventually stops forming stars;
\item the present-time accretion rates we derive are very uncertain, but even so there are indications that some large spirals may not need ongoing gas accretion.
\end{enumerate}

An obvious development of this study would be to incorporate metallicity.
Our results on the amount of gas infall needed for the Galaxy are broadly in agreement with the amount of infalling material needed by chemical evolution models. 
Our independent determination of the infall rate as a function of time and radius could be taken as an input in chemical evolution models to obtain a better understanding of the accretion mechanisms that feed the star formation in galaxy discs.

\section*{Acknowledgments}

We thank Andrew Hopkins and Donatella Romano for very helpful advise and stimulating discussions. We are grateful to an anonymous referee for helpful and constructive comments. FF was supported by the PRIN-MIUR 2008SPTACC. MT was supported for this research through a stipend from the International Max Planck Research School (IMPRS) for Astronomy and Astrophysics at the Universities of Bonn and Cologne, and by the SFB 956 "Conditions and Impact of Star Formation" by the Deutsche Forschungsgemeinschaft (DFG).


\begin{thebibliography}{}

\bibitem[\protect\citeauthoryear{{Aumer} \& {Binney}}{{Aumer} \&
  {Binney}}{2009}]{Aumer&Binney09}
{Aumer} M.,  {Binney} J.~J.,  2009, \mnras, 397, 1286

\bibitem[\protect\citeauthoryear{{Bauermeister}, {Blitz} \&
  {Ma}}{{Bauermeister} et~al.}{2010}]{Bauermeister+10}
{Bauermeister} A.,  {Blitz} L.,    {Ma} C.,  2010, \apj, 717, 323

\bibitem[\protect\citeauthoryear{{Bell}, {McIntosh}, {Katz} \&
  {Weinberg}}{{Bell} et~al.}{2003}]{Bell+03}
{Bell} E.~F.,  {McIntosh} D.~H.,  {Katz} N.,    {Weinberg} M.~D.,  2003, \apjs,
  149, 289

\bibitem[\protect\citeauthoryear{{Bigiel}, {Leroy}, {Walter}, {Brinks}, {de
  Blok}, {Madore} \& {Thornley}}{{Bigiel} et~al.}{2008}]{Bigiel08}
{Bigiel} F.,  {Leroy} A.,  {Walter} F.,  {Brinks} E.,  {de Blok} W.~J.~G.,
  {Madore} B.,    {Thornley} M.~D.,  2008, \aj, 136, 2846

\bibitem[\protect\citeauthoryear{{Binney}, {Dehnen} \& {Bertelli}}{{Binney}
  et~al.}{2000}]{Binney+00}
{Binney} J.,  {Dehnen} W.,    {Bertelli} G.,  2000, \mnras, 318, 658

\bibitem[\protect\citeauthoryear{{Binney} \& {Merrifield}}{{Binney} \&
  {Merrifield}}{1998}]{Binney&Merrifield98}
{Binney} J.,  {Merrifield} M.,  1998, {Galactic Astronomy}.
Princeton University Press

\bibitem[\protect\citeauthoryear{{Binney}, {Nipoti} \& {Fraternali}}{{Binney}
  et~al.}{2009}]{Binney+09}
{Binney} J.,  {Nipoti} C.,    {Fraternali} F.,  2009, \mnras, 397, 1804

\bibitem[\protect\citeauthoryear{{Binney} \& {Tremaine}}{{Binney} \&
  {Tremaine}}{2008}]{Binney&Tremaine08}
{Binney} J.,  {Tremaine} S.,  2008, {Galactic Dynamics: Second Edition}.
Princeton University Press

\bibitem[\protect\citeauthoryear{{Bland-Hawthorn}}{{Bland-Hawthorn}}{2009}]{Bland-Hawthorn09}
{Bland-Hawthorn} J.,  2009, in {J.~Andersen, J.~Bland-Hawthorn, \&
  B.~Nordstr{\"o}m} ed., IAU Symposium Vol.~254 of IAU Symposium, {Warm gas
  accretion onto the Galaxy}.
pp 241--254

\bibitem[\protect\citeauthoryear{{Boissier} \& {Prantzos}}{{Boissier} \&
  {Prantzos}}{1999}]{Boissier&Prantzos99}
{Boissier} S.,  {Prantzos} N.,  1999, \mnras, 307, 857

\bibitem[\protect\citeauthoryear{{Boissier}, {Prantzos}, {Boselli} \&
  {Gavazzi}}{{Boissier} et~al.}{2003}]{Boissier+03}
{Boissier} S.,  {Prantzos} N.,  {Boselli} A.,    {Gavazzi} G.,  2003, \mnras,
  346, 1215

\bibitem[\protect\citeauthoryear{{Brinchmann}, {Charlot}, {White}, {Tremonti},
  {Kauffmann}, {Heckman} \& {Brinkmann}}{{Brinchmann}
  et~al.}{2004}]{Brinchmann+04}
{Brinchmann} J.,  {Charlot} S.,  {White} S.~D.~M.,  {Tremonti} C.,  {Kauffmann}
  G.,  {Heckman} T.,    {Brinkmann} J.,  2004, \mnras, 351, 1151

\bibitem[\protect\citeauthoryear{{Cappellari}, {Emsellem}, {Krajnovi{\'c}},
  {McDermid}, {Serra}, {Alatalo}, {Blitz}, {Bois}, {Bournaud}, {Bureau},
  {Davies}, {Davis}, {de Zeeuw}, {Khochfar}, {Kuntschner} \&
  {Lablanche}}{{Cappellari} et~al.}{2011}]{Cappellari+11}
{Cappellari} M.,  {Emsellem} E.,  {Krajnovi{\'c}} D.,  {McDermid} R.~M.,
  {Serra} P.,  {Alatalo} K.,  {Blitz} L.,  {Bois} M.,  {Bournaud} F.,  {Bureau}
  M.,  {Davies} R.~L.,  {Davis} T.~A.,  {de Zeeuw} P.~T.,  {Khochfar} S.,
  {Kuntschner} H.,    {Lablanche} P.-Y.,  2011, \mnras, 416, 1680

\bibitem[\protect\citeauthoryear{{Case} \& {Bhattacharya}}{{Case} \&
  {Bhattacharya}}{1998}]{Case&Bhattacharya98}
{Case} G.~L.,  {Bhattacharya} D.,  1998, \apj, 504, 761

\bibitem[\protect\citeauthoryear{{Chabrier}}{{Chabrier}}{2003}]{Chabrier03}
{Chabrier} G.,  2003, \pasp, 115, 763

\bibitem[\protect\citeauthoryear{{Chiappini}, {Matteucci} \&
  {Gratton}}{{Chiappini} et~al.}{1997}]{Chiappini+97}
{Chiappini} C.,  {Matteucci} F.,    {Gratton} R.,  1997, \apj, 477, 765

\bibitem[\protect\citeauthoryear{{Chiappini}, {Matteucci} \&
  {Romano}}{{Chiappini} et~al.}{2001}]{Chiappini+01}
{Chiappini} C.,  {Matteucci} F.,    {Romano} D.,  2001, \apj, 554, 1044

\bibitem[\protect\citeauthoryear{{Cignoni}, {Degl'Innocenti}, {Prada Moroni} \&
  {Shore}}{{Cignoni} et~al.}{2006}]{Cignoni+06}
{Cignoni} M.,  {Degl'Innocenti} S.,  {Prada Moroni} P.~G.,    {Shore} S.~N.,
  2006, \aap, 459, 783

\bibitem[\protect\citeauthoryear{{Collins}, {Shull} \& {Giroux}}{{Collins}
  et~al.}{2009}]{Collins+09}
{Collins} J.~A.,  {Shull} J.~M.,    {Giroux} M.~L.,  2009, \apj, 705, 962

\bibitem[\protect\citeauthoryear{{Dehnen} \& {Binney}}{{Dehnen} \&
  {Binney}}{1998}]{Dehnen&Binney98}
{Dehnen} W.,  {Binney} J.,  1998, \mnras, 294, 429

\bibitem[\protect\citeauthoryear{{Dekel} \& {Birnboim}}{{Dekel} \&
  {Birnboim}}{2006}]{Dekel&Birnboim06}
{Dekel} A.,  {Birnboim} Y.,  2006, \mnras, 368, 2

\bibitem[\protect\citeauthoryear{{Diehl}, {Halloin}, {Kretschmer}, {Lichti},
  {Sch{\"o}nfelder}, {Strong}, {von Kienlin}, {Wang}, {Jean}, {Kn{\"o}dlseder},
  {Roques}, {Weidenspointner}, {Schanne}, {Hartmann}, {Winkler} \&
  {Wunderer}}{{Diehl} et~al.}{2006}]{Diehl+06}
{Diehl} R.,  {Halloin} H.,  {Kretschmer} K.,  {Lichti} G.~G.,
  {Sch{\"o}nfelder} V.,  {Strong} A.~W.,  {von Kienlin} A.,  {Wang} W.,  {Jean}
  P.,  {Kn{\"o}dlseder} J.,  {Roques} J.,  {Weidenspointner} G.,  {Schanne} S.,
   {Hartmann} D.~H.,  {Winkler} C.,    {Wunderer} C.,  2006, \nat, 439, 45

\bibitem[\protect\citeauthoryear{{Fraternali}}{{Fraternali}}{2009}]{Fraternali09}
{Fraternali} F.,  2009, in {J.~Andersen, J.~Bland-Hawthorn, \&
  B.~Nordstr{\"o}m} ed., IAU Symposium Vol.~254 of IAU Symposium, {New evidence
  for halo gas accretion onto disk galaxies}.
pp 255--262

\bibitem[\protect\citeauthoryear{{Fraternali} \& {Binney}}{{Fraternali} \&
  {Binney}}{2006}]{Fraternali&Binney06}
{Fraternali} F.,  {Binney} J.~J.,  2006, \mnras, 366, 449

\bibitem[\protect\citeauthoryear{{Fraternali} \& {Binney}}{{Fraternali} \&
  {Binney}}{2008}]{Fraternali&Binney08}
{Fraternali} F.,  {Binney} J.~J.,  2008, \mnras, 386, 935

\bibitem[\protect\citeauthoryear{{Gogarten}, {Dalcanton} \&
  {Williams}}{{Gogarten} et~al.}{2009}]{Gogarten+09}
{Gogarten} S.~M.,  {Dalcanton} J.~J.,    {Williams} B.~F.,  2009, in The
  Evolving ISM in the Milky Way and Nearby Galaxies {The Spatially Resolved
  Star Formation History of NGC 300}

\bibitem[\protect\citeauthoryear{{Harris} \& {Zaritsky}}{{Harris} \&
  {Zaritsky}}{2009}]{Harris+09}
{Harris} J.,  {Zaritsky} D.,  2009, \aj, 138, 1243

\bibitem[\protect\citeauthoryear{{Haywood}}{{Haywood}}{2001}]{Haywood01}
{Haywood} M.,  2001, \mnras, 325, 1365

\bibitem[\protect\citeauthoryear{{Hopkins} \& {Beacom}}{{Hopkins} \&
  {Beacom}}{2006}]{Hopkins&Beacom06}
{Hopkins} A.~M.,  {Beacom} J.~F.,  2006, \apj, 651, 142

\bibitem[\protect\citeauthoryear{{Hopkins}, {Irwin} \& {Connolly}}{{Hopkins}
  et~al.}{2001}]{Hopkins+01}
{Hopkins} A.~M.,  {Irwin} M.~J.,    {Connolly} A.~J.,  2001, \apjl, 558, L31

\bibitem[\protect\citeauthoryear{{Hopkins}, {McClure-Griffiths} \&
  {Gaensler}}{{Hopkins} et~al.}{2008}]{Hopkins+08}
{Hopkins} A.~M.,  {McClure-Griffiths} N.~M.,    {Gaensler} B.~M.,  2008, \apjl,
  682, L13

\bibitem[\protect\citeauthoryear{{Jorgenson}, {Wolfe}, {Prochaska} \&
  {Carswell}}{{Jorgenson} et~al.}{2009}]{Prochaska+09}
{Jorgenson} R.~A.,  {Wolfe} A.~M.,  {Prochaska} J.~X.,    {Carswell} R.~F.,
  2009, \apj, 704, 247

\bibitem[\protect\citeauthoryear{{Joung}, {Bryan} \& {Putman}}{{Joung}
  et~al.}{2011}]{Joung+11}
{Joung} M.~R.,  {Bryan} G.~L.,    {Putman} M.~E.,  2011, ArXiv e-prints

\bibitem[\protect\citeauthoryear{{Kalberla} \& {Dedes}}{{Kalberla} \&
  {Dedes}}{2008}]{Kalberla&Dedes08}
{Kalberla} P.~M.~W.,  {Dedes} L.,  2008, \aap, 487, 951

\bibitem[\protect\citeauthoryear{{Kaufmann}, {Bullock}, {Maller}, {Fang} \&
  {Wadsley}}{{Kaufmann} et~al.}{2009}]{Kaufmann+09}
{Kaufmann} T.,  {Bullock} J.~S.,  {Maller} A.~H.,  {Fang} T.,    {Wadsley} J.,
  2009, \mnras, 396, 191

\bibitem[\protect\citeauthoryear{{Kaufmann}, {Mayer}, {Wadsley}, {Stadel} \&
  {Moore}}{{Kaufmann} et~al.}{2006}]{Kaufmann+06}
{Kaufmann} T.,  {Mayer} L.,  {Wadsley} J.,  {Stadel} J.,    {Moore} B.,  2006,
  \mnras, 370, 1612

\bibitem[\protect\citeauthoryear{{Kennicutt}
  Jr.}{{Kennicutt}}{1983}]{Kennicutt83}
{Kennicutt} Jr. R.~C.,  1983, \apj, 272, 54

\bibitem[\protect\citeauthoryear{{Kennicutt}
  Jr.}{{Kennicutt}}{1998a}]{Kennicutt98b}
{Kennicutt} Jr. R.~C.,  1998a, \araa, 36, 189

\bibitem[\protect\citeauthoryear{{Kennicutt}
  Jr.}{{Kennicutt}}{1998b}]{Kennicutt98a}
{Kennicutt} Jr. R.~C.,  1998b, \apj, 498, 541

\bibitem[\protect\citeauthoryear{{Kennicutt} Jr., {Calzetti}, {Walter},
  {Helou}, {Hollenbach}, {Armus}, {Bendo}, {Dale}, {Draine}, {Engelbracht},
  {Gordon}, {Prescott}, W., {Thornley}, {Bot}, {Brinks}, {de Blok} \& {et
  al.}}{{Kennicutt} et~al.}{2007}]{Kennicutt+07}
{Kennicutt} Jr. R.~C.,  {Calzetti} D.,  {Walter} F.,  {Helou} G.,  {Hollenbach}
  D.~J.,  {Armus} L.,  {Bendo} G.,  {Dale} D.~A.,  {Draine} B.~T.,
  {Engelbracht} C.~W.,  {Gordon} K.~D.,  {Prescott} M.~K.~M.,  W. R.~M.,
  {Thornley} M.~D.,  {Bot} C.,  {Brinks} E.,  {de Blok} E.,    {et al.} 2007,
  \apj, 671, 333

\bibitem[\protect\citeauthoryear{{Kennicutt} Jr., {Tamblyn} \&
  {Congdon}}{{Kennicutt} et~al.}{1994}]{Kennicutt+94}
{Kennicutt} Jr. R.~C.,  {Tamblyn} P.,    {Congdon} C.~E.,  1994, \apj, 435, 22

\bibitem[\protect\citeauthoryear{{Kroupa}, {Tout} \& {Gilmore}}{{Kroupa}
  et~al.}{1993}]{Kroupa+93}
{Kroupa} P.,  {Tout} C.~A.,    {Gilmore} G.,  1993, \mnras, 262, 545

\bibitem[\protect\citeauthoryear{{Krumholz} \& {Dekel}}{{Krumholz} \&
  {Dekel}}{2011}]{Krumholz&Dekel11}
{Krumholz} M.~R.,  {Dekel} A.,  2011, ArXiv e-prints

\bibitem[\protect\citeauthoryear{{Krumholz}, {Dekel} \& {McKee}}{{Krumholz}
  et~al.}{2012}]{Krumholz+12}
{Krumholz} M.~R.,  {Dekel} A.,    {McKee} C.~F.,  2012, \apj, 745, 69

\bibitem[\protect\citeauthoryear{{Krumholz}, {Leroy} \& {McKee}}{{Krumholz}
  et~al.}{2011}]{Krumholz+11}
{Krumholz} M.~R.,  {Leroy} A.~K.,    {McKee} C.~F.,  2011, \apj, 731, 25

\bibitem[\protect\citeauthoryear{{Lah}, {Chengalur}, {Briggs}, {Colless}, {de
  Propris}, {Pracy}, {de Blok}, {Fujita}, {Ajiki}, {Shioya}, {Nagao},
  {Murayama}, {Taniguchi}, {Yagi} \& {Okamura}}{{Lah} et~al.}{2007}]{Lah+07}
{Lah} P.,  {Chengalur} J.~N.,  {Briggs} F.~H.,  {Colless} M.,  {de Propris} R.,
   {Pracy} M.~B.,  {de Blok} W.~J.~G.,  {Fujita} S.~S.,  {Ajiki} M.,  {Shioya}
  Y.,  {Nagao} T.,  {Murayama} T.,  {Taniguchi} Y.,  {Yagi} M.,    {Okamura}
  S.,  2007, \mnras, 376, 1357

\bibitem[\protect\citeauthoryear{{Lehner} \& {Howk}}{{Lehner} \&
  {Howk}}{2011}]{Lehner&Howk11}
{Lehner} N.,  {Howk} J.~C.,  2011, Science, 334, 955

\bibitem[\protect\citeauthoryear{{Leroy}, {Walter}, {Brinks}, {Bigiel}, {de
  Blok}, {Madore} \& {Thornley}}{{Leroy} et~al.}{2008}]{Leroy+08}
{Leroy} A.~K.,  {Walter} F.,  {Brinks} E.,  {Bigiel} F.,  {de Blok} W.~J.~G.,
  {Madore} B.,    {Thornley} M.~D.,  2008, \aj, 136, 2782

\bibitem[\protect\citeauthoryear{{Lyne}, {Manchester} \& {Taylor}}{{Lyne}
  et~al.}{1985}]{Lyne+85}
{Lyne} A.~G.,  {Manchester} R.~N.,    {Taylor} J.~H.,  1985, \mnras, 213, 613

\bibitem[\protect\citeauthoryear{{Maeder} \& {Meynet}}{{Maeder} \&
  {Meynet}}{1989}]{Maeder&Meynet89}
{Maeder} A.,  {Meynet} G.,  1989, \aap, 210, 155

\bibitem[\protect\citeauthoryear{{Maller} \& {Bullock}}{{Maller} \&
  {Bullock}}{2004}]{Maller&Bullock04}
{Maller} A.~H.,  {Bullock} J.~S.,  2004, \mnras, 355, 694

\bibitem[\protect\citeauthoryear{{Marasco}, {Fraternali} \& {Binney}}{{Marasco}
  et~al.}{2011}]{Marasco+11}
{Marasco} A.,  {Fraternali} F.,    {Binney} J.~J.,  2011, ArXiv e-prints

\bibitem[\protect\citeauthoryear{{Marinacci}, {Binney}, {Fraternali}, {Nipoti},
  {Ciotti} \& {Londrillo}}{{Marinacci} et~al.}{2010}]{Marinacci+10}
{Marinacci} F.,  {Binney} J.,  {Fraternali} F.,  {Nipoti} C.,  {Ciotti} L.,
  {Londrillo} P.,  2010, \mnras, 404, 1464

\bibitem[\protect\citeauthoryear{{Marinacci}, {Fraternali}, {Nipoti}, {Binney},
  {Ciotti} \& {Londrillo}}{{Marinacci} et~al.}{2011}]{Marinacci+11}
{Marinacci} F.,  {Fraternali} F.,  {Nipoti} C.,  {Binney} J.,  {Ciotti} L.,
  {Londrillo} P.,  2011, \mnras, 415, 1534

\bibitem[\protect\citeauthoryear{{Matteucci}, {Romano} \& {Molaro}}{{Matteucci}
  et~al.}{1999}]{Matteucci+99}
{Matteucci} F.,  {Romano} D.,    {Molaro} P.,  1999, \aap, 341, 458

\bibitem[\protect\citeauthoryear{{Matteucci, F.}}{{Matteucci,
  F.}}{2001}]{Matteucci01}
{Matteucci, F.} ed. 2001, {The chemical evolution of the Galaxy} Vol.~253 of
  Astrophysics and Space Science Library

\bibitem[\protect\citeauthoryear{{Minchev}, {Famaey}, {Combes}, {Di Matteo},
  {Mouhcine} \& {Wozniak}}{{Minchev} et~al.}{2011}]{Minchev+11}
{Minchev} I.,  {Famaey} B.,  {Combes} F.,  {Di Matteo} P.,  {Mouhcine} M.,
  {Wozniak} H.,  2011, \aap, 527, A147

\bibitem[\protect\citeauthoryear{{Murray} \& {Rahman}}{{Murray} \&
  {Rahman}}{2010}]{Murray&Rahman10}
{Murray} N.,  {Rahman} M.,  2010, \apj, 709, 424

\bibitem[\protect\citeauthoryear{{Naab} \& {Ostriker}}{{Naab} \&
  {Ostriker}}{2006}]{Naab&Ostriker06}
{Naab} T.,  {Ostriker} J.~P.,  2006, \mnras, 366, 899

\bibitem[\protect\citeauthoryear{{Nakanishi} \& {Sofue}}{{Nakanishi} \&
  {Sofue}}{2006}]{Nakanishi&Sofue06}
{Nakanishi} H.,  {Sofue} Y.,  2006, \pasj, 58, 847

\bibitem[\protect\citeauthoryear{{Nipoti}}{{Nipoti}}{2010}]{Nipoti10}
{Nipoti} C.,  2010, \mnras, 406, 247

\bibitem[\protect\citeauthoryear{{Pagel} \& {Patchett}}{{Pagel} \&
  {Patchett}}{1975}]{Pagel&Patchett75}
{Pagel} B.~E.~J.,  {Patchett} B.~E.,  1975, \mnras, 172, 13

\bibitem[\protect\citeauthoryear{{Panter}, {Jimenez}, {Heavens} \&
  {Charlot}}{{Panter} et~al.}{2007}]{Panter+07}
{Panter} B.,  {Jimenez} R.,  {Heavens} A.~F.,    {Charlot} S.,  2007, \mnras,
  378, 1550

\bibitem[\protect\citeauthoryear{{Peek}}{{Peek}}{2009}]{Peek09}
{Peek} J.~E.~G.,  2009, \apj, 698, 1429

\bibitem[\protect\citeauthoryear{{Peek}, {Putman} \& {Sommer-Larsen}}{{Peek}
  et~al.}{2008}]{Peek+08}
{Peek} J.~E.~G.,  {Putman} M.~E.,    {Sommer-Larsen} J.,  2008, \apj, 674, 227

\bibitem[\protect\citeauthoryear{{Piontek} \& {Steinmetz}}{{Piontek} \&
  {Steinmetz}}{2011}]{Piontek&Steinmetz11}
{Piontek} F.,  {Steinmetz} M.,  2011, \mnras, 410, 2625

\bibitem[\protect\citeauthoryear{{Prantzos}}{{Prantzos}}{2008}]{Prantzos08}
{Prantzos} N.,  2008, in {Charbonnel} C.,  {Zahn} J.-P.,  eds, EAS Publications
  Series Vol.~32 of EAS Publications Series, {An Introduction to Galactic
  Chemical Evolution}.
Cambridge University Press, pp 311--356

\bibitem[\protect\citeauthoryear{{Rocha-Pinto}, {Scalo}, {Maciel} \&
  {Flynn}}{{Rocha-Pinto} et~al.}{2000}]{Rocha-Pinto+00}
{Rocha-Pinto} H.~J.,  {Scalo} J.,  {Maciel} W.~J.,    {Flynn} C.,  2000, \aap,
  358, 869

\bibitem[\protect\citeauthoryear{{Ro{\v s}kar}, {Debattista}, {Quinn},
  {Stinson} \& {Wadsley}}{{Ro{\v s}kar} et~al.}{2008}]{Roskar+08}
{Ro{\v s}kar} R.,  {Debattista} V.~P.,  {Quinn} T.~R.,  {Stinson} G.~S.,
  {Wadsley} J.,  2008, \apjl, 684, L79

\bibitem[\protect\citeauthoryear{{Roychowdhury}, {Chengalur}, {Begum} \&
  {Karachentsev}}{{Roychowdhury} et~al.}{2009}]{Roychowdhury+09}
{Roychowdhury} S.,  {Chengalur} J.~N.,  {Begum} A.,    {Karachentsev} I.~D.,
  2009, \mnras, 397, 1435

\bibitem[\protect\citeauthoryear{{Sancisi}, {Fraternali}, {Oosterloo} \& {van
  der Hulst}}{{Sancisi} et~al.}{2008}]{Sancisi+08}
{Sancisi} R.,  {Fraternali} F.,  {Oosterloo} T.,    {van der Hulst} T.,  2008,
  \aapr, 15, 189

\bibitem[\protect\citeauthoryear{{Scalo}}{{Scalo}}{1986}]{Scalo86}
{Scalo} J.~M.,  1986, \fcp, 11, 1

\bibitem[\protect\citeauthoryear{{Schaye}}{{Schaye}}{2004}]{Schaye04}
{Schaye} J.,  2004, \apj, 609, 667

\bibitem[\protect\citeauthoryear{{Schmidt}}{{Schmidt}}{1959}]{Schmidt59}
{Schmidt} M.,  1959, \apj, 129, 243

\bibitem[\protect\citeauthoryear{{Searle} \& {Sargent}}{{Searle} \&
  {Sargent}}{1972}]{Searle&Sargent72}
{Searle} L.,  {Sargent} W.~L.~W.,  1972, Comments on Astrophysics and Space
  Physics, 4, 59

\bibitem[\protect\citeauthoryear{{Sellwood} \& {Binney}}{{Sellwood} \&
  {Binney}}{2002}]{Sellwood&Binney02}
{Sellwood} J.~A.,  {Binney} J.~J.,  2002, \mnras, 336, 785

\bibitem[\protect\citeauthoryear{{Thom}, {Peek}, {Putman}, {Heiles}, {Peek} \&
  {Wilhelm}}{{Thom} et~al.}{2008}]{Thom+08}
{Thom} C.,  {Peek} J.~E.~G.,  {Putman} M.~E.,  {Heiles} C.,  {Peek} K.~M.~G.,
   {Wilhelm} R.,  2008, \apj, 684, 364

\bibitem[\protect\citeauthoryear{{Tinsley}}{{Tinsley}}{1980}]{Tinsley80}
{Tinsley} B.~M.,  1980, \fcp, 5, 287

\bibitem[\protect\citeauthoryear{{Tinsley} \& {Larson}}{{Tinsley} \&
  {Larson}}{1978}]{Tinsley&Larson78}
{Tinsley} B.~M.,  {Larson} R.~B.,  1978, \apj, 221, 554

\bibitem[\protect\citeauthoryear{{Tosi}}{{Tosi}}{1988a}]{Tosi88a}
{Tosi} M.,  1988a, \aap, 197, 33

\bibitem[\protect\citeauthoryear{{Tosi}}{{Tosi}}{1988b}]{Tosi88b}
{Tosi} M.,  1988b, \aap, 197, 47

\bibitem[\protect\citeauthoryear{{Twarog}}{{Twarog}}{1980}]{Twarog80}
{Twarog} B.~A.,  1980, \apj, 242, 242

\bibitem[\protect\citeauthoryear{{van de Voort}, {Schaye}, {Booth} \& {Dalla
  Vecchia}}{{van de Voort} et~al.}{2011}]{vdVoort+11}
{van de Voort} F.,  {Schaye} J.,  {Booth} C.~M.,    {Dalla Vecchia} C.,  2011,
  \mnras, 415, 2782

\bibitem[\protect\citeauthoryear{{Vincoletto}, {Matteucci}, {Calura}, {Silva}
  \& {Granato}}{{Vincoletto} et~al.}{2012}]{Vincoletto+12}
{Vincoletto} L.,  {Matteucci} F.,  {Calura} F.,  {Silva} L.,    {Granato} G.,
  2012, \mnras, 421, 3116

\bibitem[\protect\citeauthoryear{{Walter}, {Brinks}, {de Blok}, {Bigiel},
  {Kennicutt} Jr., {Thornley} \& {Leroy}}{{Walter} et~al.}{2008}]{Walter+08}
{Walter} F.,  {Brinks} E.,  {de Blok} W.~J.~G.,  {Bigiel} F.,  {Kennicutt} Jr.
  R.~C.,  {Thornley} M.~D.,    {Leroy} A.,  2008, \aj, 136, 2563

\bibitem[\protect\citeauthoryear{{Weisz}, {Dalcanton}, {Williams}, {Gilbert},
  {Skillman}, {Seth}, {Dolphin}, {McQuinn}, {Gogarten}, {Holtzman}, {Rosema},
  {Cole}, {Karachentsev} \& {Zaritsky}}{{Weisz} et~al.}{2011}]{Weisz+11}
{Weisz} D.~R.,  {Dalcanton} J.~J.,  {Williams} B.~F.,  {Gilbert} K.~M.,
  {Skillman} E.~D.,  {Seth} A.~C.,  {Dolphin} A.~E.,  {McQuinn} K.~B.~W.,
  {Gogarten} S.~M.,  {Holtzman} J.,  {Rosema} K.,  {Cole} A.,  {Karachentsev}
  I.~D.,    {Zaritsky} D.,  2011, ArXiv e-prints

\bibitem[\protect\citeauthoryear{{Wyder}, {Martin}, {Barlow}, {Foster},
  {Friedman}, {Morrissey}, {Neff}, {Neill}, {Schiminovich}, {Seibert},
  {Bianchi}, {Donas}, {Heckman}, {Lee}, {Madore}, {Milliard}, {Rich}, {Szalay}
  \& {Yi}}{{Wyder} et~al.}{2009}]{Wyder+09}
{Wyder} T.~K.,  {Martin} D.~C.,  {Barlow} T.~A.,  {Foster} K.,  {Friedman}
  P.~G.,  {Morrissey} P.,  {Neff} S.~G.,  {Neill} J.~D.,  {Schiminovich} D.,
  {Seibert} M.,  {Bianchi} L.,  {Donas} J.,  {Heckman} T.~M.,  {Lee} Y.-W.,
  {Madore} B.~F.,  {Milliard} B.,  {Rich} R.~M.,  {Szalay} A.~S.,    {Yi}
  S.~K.,  2009, \apj, 696, 1834

\bibitem[\protect\citeauthoryear{{Zwaan}, {van der Hulst}, {Briggs},
  {Verheijen} \& {Ryan-Weber}}{{Zwaan} et~al.}{2005}]{Zwaan+05}
{Zwaan} M.~A.,  {van der Hulst} J.~M.,  {Briggs} F.~H.,  {Verheijen} M.~A.~W.,
    {Ryan-Weber} E.~V.,  2005, \mnras, 364, 1467

\end{thebibliography}

\appendix

\bsp

\end{document}